\documentclass[useAMS,usenatbib,usegraphicx,onecolumn]{mn2e}
\def\be{\begin{equation}}
\def\ee{\end{equation}}
\def\simlt{\mathrel{\hbox{\rlap{\hbox{\lower3pt\hbox{$\sim$}}}\hbox{$<$}}}}
\def\simgt{\mathrel{\hbox{\rlap{\hbox{\lower3pt\hbox{$\sim$}}}\hbox{$>$}}}}

\def\Aearly{^{A_\bullet, {\rm early}}}
\def\AQSO{^{A_\bullet, {\rm QSO}}}
\def\Edd{_{\rm Edd}}
\def\Hubble{_{\rm Hubble}}
\def\QSO{^{\rm QSO}}

\def\acc{_{\rm acc}}

\def\bh{_\bullet}
\def\bhaccr{_{\bullet,\rm acc}}
\def\bhL{_{\bullet,L}}

\def\bol{_{\rm bol}}
\def\d{{\rm d}}
\def\e{_{\rm e}}
\def\early{^{\rm early}}
\def\ir{_{\rm ir}}
\def\local{_{\rm local}}
\def\merge{_{\rm merge}}

\def\qso{_{\rm QSO}}

\def\dex{{\rm\,dex}}
\def\ergs{{\rm\,erg\,s^{-1}}}
\def\kms{{\rm\,km\,s^{-1}}}
\def\kpc{{\rm\,kpc}}
\def\mag{{\rm\,mag}}
\def\msun{{\rm\,M_\odot}}

\def\Lgsun{{\rm\,L_{g^*,\odot}}}

\def\pc{{\rm\,pc}}
\def\yr{{\rm\,yr}}

\def\Mpc{{\rm\,Mpc}}

\title[Growth of massive black holes]{Observational constraints on growth of massive black holes}
\author[Q. Yu \& S. Tremaine]{Qingjuan Yu\thanks{yqj@astro.princeton.edu} and Scott Tremaine\thanks{tremaine@astro.princeton.edu}\\
Princeton University Observatory, Peyton Hall, Princeton, NJ~08544-1001, USA
}

\begin{document}

\label{firstpage}
\maketitle

\begin{abstract}
\noindent
We study the observational constraints on the growth of massive black holes
(BHs) in galactic nuclei.
We use the velocity dispersions of early-type galaxies obtained by the Sloan
Digital Sky Survey and the relation between BH mass and velocity dispersion
to estimate the local BH mass density to be
$\rho\bh(z=0)\simeq (2.5\pm0.4)\times 10^5 h_{0.65}^2\msun\Mpc^{-3}$.
We also use the QSO luminosity function from the 2dF Redshift Survey
to estimate the BH mass density accreted during optically bright QSO phases.
The local BH mass density is consistent with the density
accreted during optically bright QSO phases if QSOs have a
mass-to-energy conversion efficiency $\epsilon\simeq 0.1$.
By studying the continuity equation for the BH mass distribution,
including the effect of BH mergers, we find relations
between the local BH mass function and the QSO luminosity function.
If the BH mass is assumed to be conserved during
BH mergers, comparison of the predicted relations with the observations
suggests that luminous QSOs ($L\bol\ga 10^{46}\ergs$)
have a high efficiency (e.g. $\epsilon\sim$0.2, which is possible for
thin-disk accretion onto a Kerr BH) and
the growth of high-mass BHs ($\ga 10^8\msun$) comes mainly
from accretion during optically bright QSO phases, or
that luminous QSOs have a super-Eddington luminosity.
If luminous QSOs are not accreting with super-Eddington luminosities and
the growth of low-mass BHs also occurs mainly during optically bright
QSO phases, less luminous QSOs must accrete with a low efficiency $<0.1$;
alternatively, they may accrete with high efficiency, but a significant
fraction should be obscured.
We estimate that the mean lifetime of luminous QSOs
($L\bol\ga 10^{46}\ergs$) is (3--13)$\times 10^7\yr$,
which is comparable to the Salpeter time.
We also investigate the case in which total BH mass decreases during
BH mergers due to gravitational radiation; in the extreme case in which
total BH entropy is conserved,
the observations again suggest that BHs in most luminous QSOs
are Kerr BHs accreting with an efficiency $\ga 0.1$.
\end{abstract}


\section{Introduction} \label{sec:intro}

\noindent
Most nearby galaxies contain massive dark objects at their centers
(e.g. Kormendy \& Richstone 1995, Kormendy \& Gebhardt 2001),
which are presumably black holes (BHs).
The existence of these objects was predicted by arguments based on
quasi-stellar object (QSO) energetics and demography (e.g. So{\l}tan 1982,
Rees 1984).
Studies of central BHs in nearby galaxies have also revealed a
tight correlation between BH mass and galactic velocity dispersion
\citep{FM00,Gebhardt00}, and a less tight correlation between BH mass and
the luminosity (or mass) of the hot stellar component of the host galaxy
(e.g. Kormendy \& Gebhardt 2001 and references therein;
by ``hot''component we mean either an elliptical galaxy or the bulge of
a spiral or S0 galaxy).
These correlations strongly suggest a close link between the formation and
evolution of galaxies and their central BHs.

The simplest and most elegant argument relating the properties of QSOs to
BHs in nearby galaxies is due to \citet{S82}, who pointed out that
the luminosity function of QSOs as a function of redshift traces the
accretion history of these BHs.
The BH mass density due to bright QSO phases can be estimated directly
from the local energy density in QSO photons and an assumed mass-to-energy
conversion efficiency
(So{\l}tan 1982; Chokshi \& Turner 1992).
Comparison of this density with the local BH mass density provides
considerable insight into the formation and growth of massive BHs
(e.g. \citealt{HK01} and references therein).

In \S~\ref{sec:bhden}, we provide new estimates of the local BH
mass density and the BH mass density accreted during bright QSO phases.
There are several reasons why a re-analysis of this classic problem is timely.
(i) The past several years have seen dramatic improvements in optical QSO
surveys, both in terms of the total numbers of QSOs and the parameter ranges
(i.e., luminosities and redshifts) over which the QSO luminosity function
is reliably determined.
For example, the 2dF QSO redshift survey \citep{Boyle00} and the Large Bright
QSO Survey \citep{HFC95} have found $\sim 6000$ QSOs
with absolute magnitude $-26<M_B<-23$
($\Omega_{\rm m}=1, \Omega_\Lambda=0, h=0.5$), at redshifts $0.35<z<2.3$;
the Sloan Digital Sky Survey (SDSS) has found several hundred QSOs
with redshifts $3.6<z<5.0$ (e.g. \citealt{Fan01}; see also other surveys
at high redshifts, e.g. Schmidt, Schneider \& Gunn 1995 etc.)
(ii) The SDSS has provided a sample of $\sim 9000$ nearby early-type
galaxies (elliptical galaxies and S0 galaxies; Bernardi et al. 2001)
with accurate luminosities and velocity dispersions.
Using the tight correlation between BH mass and galactic velocity dispersion,
we can estimate the BH masses in these galaxies and
thereby obtain the mass function of BHs at $z=0$.
(iii) A large number of obscured active galactic nuclei (AGNs)
(e.g. $\sim$4--10 times the number of unobscured AGNs) is required to explain
the observed X-ray background spectrum (e.g. \citealt{FI99,GSH01}).
The BH mass density inferred from the X-ray background by \citet{FI99}
is larger than the one derived from optically bright QSOs by
\citet{CT92} by a factor of 3--4.
BHs in local galaxies are the remnants of both obscured and
unobscured AGNs.
Thus, an accurate estimate of the local BH mass density and
the accreted mass density from optically bright QSOs can also
constrain the fraction of obscured AGNs for comparison to models
of the X-ray background.

If BH growth occurs mainly during optically bright QSO phases, then not only
should the BH mass density in local galaxies be consistent with the total
energy density in QSO light, but also the BH mass distribution in local
galaxies should be consistent with the QSO luminosity function
(e.g. \citealt{SB92,CHO01,MS01}).
In \S~\ref{sec:clue}, by studying the continuity equation for the BH mass
distribution, we test this hypothesis and give more general observational
constraints (e.g. on the mass-to-energy conversion efficiency and lifetime
of QSOs) on the growth of massive BHs in galactic nuclei.
The effects of BH mergers (caused by galaxy mergers) on the BH mass
distribution are also considered.
Since the physics of BH mergers is poorly understood, we consider
two extreme cases: one is the classical case
(mergers do not emit gravitational radiation and the total BH mass
is conserved during BH mergers), which is implicitly assumed
in So{\l}tan's argument; the other is the adiabatic case
(total BH area and entropy is conserved during BH mergers and mass is radiated
away as gravitational radiation; see also \citealt{Cv01}).
Our conclusions are summarized in \S~\ref{sec:discon}.

In this paper, the Hubble constant is written as $H_0=100h\kms\Mpc^{-1}$;
and if not otherwise specified, the cosmological model used is
$(\Omega_{\rm M},\Omega_{\Lambda},h)=(0.3,0.7,0.65)$ \citep{Wang00}.

%%%%%%%%%%%%%%%%%%%%%%%%%%%%%%%%%%%%%%%%%%%%%%%%%%%%%%%%%%%%%%%%%%%%%%%%%%%
\section{Total BH mass density} \label{sec:bhden}

%==========================================================================
\subsection{Early-type galaxies in the SDSS} \label{subsec:earlytype}

\noindent
The SDSS will image $\sim 10^4$ square degrees of the sky
in five bands $(u, g, r, i, z)$ with central wavelengths
(3560, 4680, 6180, 7500, 8870)\AA, and take spectra
of $\sim10^6$ galaxies and $\sim10^5$ QSOs \citep{York00}.
Among the spectra to be taken by SDSS, there will be roughly
$2\times10^5$ spectra of early-type galaxies.
A sample of nearly 9000 of these with well-measured velocity
dispersions ($\delta\log\sigma\sim$0.02--$0.06\dex$ with a median value
$\sim0.03\dex$), in the redshift range $0.01\le z\le 0.3$, has been selected
from early SDSS observations by \citet{Bernardi01}.
The joint distribution of luminosities, sizes and velocity dispersions for
this sample is well described by a tri-variate Gaussian distribution
in the variables $M=-2.5\log L$, $R=\log R_o$, and $V=\log\sigma$,
where $L$ is the total luminosity,
$R_o$ is the effective radius and $\sigma$ is the line-of-sight
velocity dispersion within a circular aperture of radius $R_o/8$.
A maximum-likelihood fit to the sample gives the mean values of the above
variables, $\langle M\rangle=-19.65+5\log h-1.15z$
(the $1.15z$ term represents evolution),
$\langle R\rangle=0.36$, and $\langle V\rangle =2.20$ in the $g^*$ band
(here $10^R$ and $10^V$ are in units of $h^{-1}\kpc$ and $\kms$, respectively;
$g^*$ is used rather than $g$, similarly for other bands, because the
photometric calibration is preliminary; for details, see Stoughton et al. 2002)
and their standard deviations
$\sigma_M=0.84$, $\sigma_R=0.25$ and $\sigma_V=0.11$
(cf. table 2 in Bernardi et al. 2001).
By extrapolating this luminosity distribution at both faint and bright ends,
the average comoving number density of early-type galaxies
is found to be
\be
\phi_*=(5.8\pm0.3)\times 10^{-3}h^3\Mpc^{-3}.
\label{eq:numden}
\ee
For comparison to other studies, we must convert magnitudes in the SDSS
$g^*$ band to the Johnson-Morgan $B$ band.
We take the color transformation to be:
\be
B=g^*+0.47(g^*-r^*)+0.17,
\label{eq:Bgtrans}
\ee
which is derived from the preliminary SDSS calibration \citep{Smith01}
of the Landolt stars \citep{L92a,L92b}.
The standard deviation in this relation is $0.03\mag$.
The mean color of the early-type galaxies in Bernardi et al.'s SDSS sample
is $g^*-r^*\simeq0.74-0.30z$ with standard deviation $0.06\mag$ in the
colors of individual galaxies (cf. table 6 in Bernardi et al. 2001).

%==========================================================================
\subsection{Local BH mass density} \label{subsec:bhdenlocal}

\noindent
The SDSS data provide us with the distributions of velocity
dispersion and luminosity in nearby early-type galaxies.
We denote the luminosity function and the velocity-dispersion
distribution of early-type galaxies as $n_L\early(L,z)$ and
$n_\sigma\early(\sigma,z)$, where $n_L\early(L,z)\d L$
[or $n_\sigma\early(\sigma,z)\d\sigma$] represents the comoving number density
of early-type galaxies in the range $L\rightarrow L+\d L$
(or $\sigma\rightarrow\sigma+\d\sigma$) at redshift $z$.
The distributions $n_L\early(L,z)$ and $n_\sigma\early(\sigma,z)$
are obtained by integrating the tri-variate distribution in
$M=-2.5\log L$, $R=\log R_o$ and $V=\log\sigma$ over two of these variables.
Thus, the average number density of these galaxies is
(eq.~\ref{eq:numden})
\be
\phi_*=\int_0^\infty n_L\early(L,z=0)\d L
=\int_0^\infty n_\sigma\early(\sigma,z=0)\d\sigma,
\ee
and the luminosity density is
\be
j=\int_0^\infty L n_L^{\rm early}(L,z=0)dL.
\ee
The luminosity density derived from the Bernardi et al. sample is
$j_{g^*}=6.1\times 10^7 h\, \Lgsun\Mpc^{-3}$. For comparison, the total
luminosity density derived from SDSS is $j_{g^*}=(2.8\pm0.4)\times 10^8 h\,
\Lgsun \Mpc^{-3}$ (Blanton et al. 2001). This total luminosity density is a
factor of 1.5 to 2 times higher than earlier estimates, but Blanton et
al. argue that the difference arises because earlier surveys neglect low
surface-brightness regions that are included in SDSS.
The luminosity density in the Blanton et al. sample can be compared directly
to the Bernardi et al. sample, as follows. Strateva et al. (2001) show that
early- and late-type galaxies in SDSS can be separated by a cut in
concentration index at $C=2.6$; applying this cut to the Blanton et al. sample
we derive $j_{g^*}=7.4\times 10^7 h\, \Lgsun\Mpc^{-3}$ for early-type
galaxies. This result differs by only 20\% from the estimate of Bernardi et
al., which uses several parameters including the concentration index to
separate early- from late-type galaxies. Thus the normalization of the sample
isolated by Bernardi et al. appears to be consistent with other early-type
galaxy samples in SDSS.

The relation between BH mass and galactic velocity dispersion is
\citep{Tremaine02}:
\be
M\bh=(1.5\pm 0.2)\times 10^8\msun
\left(\frac{\sigma\e}{200\kms}\right)^{\alpha_1},
\label{eq:msigma}
\ee
where $\alpha_1=4.02\pm0.32$, $M\bh$ is the BH mass, and $\sigma\e$ is
the luminosity-weighted line-of-sight
velocity dispersion within a slit extending to the effective radius.
Note that $\sigma\e$ in equation (\ref{eq:msigma}) is the velocity
dispersion within a slit extending to the effective radius $R_o$
\citep{Gebhardt00} while $\sigma$ in the SDSS is the
velocity dispersion within a circular aperture extending to $R_o/8$; however,
replacing $\sigma\e$ with $\sigma$ in equation (\ref{eq:msigma})
will not cause much difference as the two definitions should
give very similar results (Tremaine et al. 2002).
\citet{MF01a} give an alternative version of the correlation (\ref{eq:msigma})
that is based on the velocity dispersion within $R_o/8$, and find a steeper
slope, $\alpha_1=4.72$. The reasons for the difference in slopes are discussed
by Tremaine et al. (2002). 
We have performed all of the calculations below using both versions of the
correlation and the differences are negligible.

The mean relation between the $B$-band luminosity of the hot stellar
component and the BH mass is \citep{KG01}
\be
M\bh=0.7\times10^8\msun\left(\frac{L_{B,\rm
hot}}{10^{10}L_{B_\odot}}\right)^{\alpha_2}, 
\label{eq:mlb}
\ee
where $\alpha_2=1.08$ \citep{KG01}. The scatter in relation (\ref{eq:mlb}) is
substantially larger than the scatter in relation (\ref{eq:msigma}).
The statistical uncertainties in $\alpha_2$ and in the
normalization are about $\pm0.15$--0.2 and 30\%, respectively. A similar
relation to equation (\ref{eq:mlb}) was derived by Magorrian et al. (1998),
but with normalization larger by a factor of 3--4, presumably because of the
limitations of ground-based spectroscopy and isotropic dynamical models.

Most of the galaxies in \citet{Tremaine02} and \citet{KG01} have distances
derived from Tonry et al. (2001), which implies a Hubble constant
$h=0.73\pm0.04\pm0.11$ \citep{Blakeslee02}, and the normalizations in equations
(\ref{eq:msigma}) and (\ref{eq:mlb}) have been adjusted to our assumed
Hubble constant $h=0.65$.

Because the mass--dispersion relation is the tighter of the two relations,
we will first use equation (\ref{eq:msigma}) to estimate the local BH mass
density.
Thus, with the distribution of velocity dispersion
$n_{\sigma}\early(\sigma,z=0)$ determined by \citet{Bernardi01},
we have the local BH mass density in early-type galaxies:
\begin{eqnarray}
\rho\early\bh(z=0) & = &\int M\bh n_{M\bh}\early(M\bh,z=0)\d M\bh=
\int M\bh n_\sigma\early(\sigma,z=0)\frac{\d \sigma}{d M\bh}\d M\bh
\nonumber \\
& = &
(1.6\pm 0.2)\times 10^5\msun\Mpc^{-3},
\label{eq:rhoearlysigma}
\end{eqnarray}
where $\sigma$ has been equated to $\sigma\e$ and $\d\sigma/\d M\bh$ is
obtained from equation (\ref{eq:msigma}).
The error estimate represents only the effects of the uncertainties in
equations (\ref{eq:numden}) and (\ref{eq:msigma}).

We may also use the mass--luminosity relation (\ref{eq:mlb}) to estimate
$\rho\bh\early$.  We note first that the luminosity provided in
\citet{Bernardi01} is the total luminosity.  Early-type galaxies include both
elliptical galaxies and S0 galaxies.  For S0 galaxies, the bulge and disk both
contribute to the total luminosity.  Therefore, when applying equation
(\ref{eq:mlb}), we need to correct the total luminosity of an S0 galaxy to its
bulge luminosity.  Since the BH mass is approximately proportional to the
luminosity of the hot stellar component (i.e., $\alpha_2=1.08$ in
eq.~\ref{eq:mlb} is approximately 1), we may simply correct the total
luminosity density of early-type galaxies in \citet{Bernardi01} by the ratio
of the luminosity density of hot components to the total luminosity density.
\citet{FHP98} estimated the relative luminosity densities $f_i$
(in the $B$ band) for elliptical galaxies, S0 galaxies, spiral and
irregular galaxies, and hot components.
They found
$(f_{\rm E},f_{\rm S0},f_{\rm Spiral+Irr})=(0.12,0.23,0.65)$ and
$f_{\rm hot}=0.385$.
They also estimate that in S0 galaxies, the bulge luminosity is
a fraction $f_{\rm bulge,S0}/f_{\rm S0}=0.64$ of the total luminosity.
Thus, the total luminosity in hot components of early-type galaxies occupies
a fraction
\be
{f_{\rm E}+f_{\rm bulge,S0}\over f_{\rm E}+f_{\rm S0}}\simeq0.76
\label{eq:lfrac}
\ee
of the total luminosity in early-type galaxies.
The local BH mass density in early-type galaxies is therefore given by:
\begin{eqnarray}
\rho\bhL\early(z=0)& = &\int M\bh n_{M\bh}\early(M\bh,z=0)\d M\bh\simeq
0.76\int M\bh n_L\early(L,z=0)\frac{\d L}{\d M\bh}\d M\bh \nonumber \\
& = &
2.0\times 10^5 \msun\Mpc^{-3},
\label{eq:rhoearlyL}
\end{eqnarray}
where the color transformation between $g^*$ and $B$
(eq.~\ref{eq:Bgtrans}) has been used.

Massive BHs exist not only in early-type galaxies but also in spiral bulges.
Hence, we must augment the estimates (\ref{eq:rhoearlysigma}) and
(\ref{eq:rhoearlyL}) by the contribution from BHs in spiral bulges.  As above,
we assume that the luminosity fraction is approximately equal to the BH mass
fraction, and thus the total BH mass density is $f_{\rm hot}/(f_{\rm E}+f_{\rm
bulge,S0})
\simeq1.44$ times the BH mass density in early-type galaxies,
i.e., from equation (\ref{eq:rhoearlysigma})
\be
\rho\bh(z=0)\simeq1.44\rho\bh\early
=(2.3\pm0.3)\times 10^5 \msun\Mpc^{-3},
\label{eq:rhosigmaa}
\ee
and from equation (\ref{eq:rhoearlyL})
\be
\rho\bhL(z=0)\simeq1.44\rho\bhL\early\simeq
2.9\times 10^5\msun\Mpc^{-3}.
\label{eq:rhola}
\ee

One additional correction to these estimates must be considered. In deriving
equations (\ref{eq:rhoearlysigma}) and (\ref{eq:rhoearlyL}), we have ignored
any intrinsic dispersion in the mass--dispersion relation and the
mass--luminosity relations (eqs.~\ref{eq:msigma} and \ref{eq:mlb}).  Note that
equations (\ref{eq:msigma}) and (\ref{eq:mlb}) are fitted in the $\log
M\bh$--$\log\sigma$ and $\log M\bh$--$\log L_{B,\rm hot}$ spaces.  Let us
assume that
the distribution in $\log M\bh$ at a given value of
$\sigma$ or $L_{B,\rm hot}$ is Gaussian, with mean given by the relations
(\ref{eq:msigma}) or (\ref{eq:mlb}), and standard deviation
$\Delta_{\log M\bh}$ that is independent of $\sigma$ or $L_{B,\rm hot}$.
Then our
estimates of the mass density must be increased by a factor
\be
\exp\left[{1\over 2}(\Delta_{\log M\bh}\ln10)^2\right]=1+{1\over 2}
(\ln10)^2\Delta^2_{\log M\bh}+\hbox{O}(\Delta^4_{\log M\bh}).
\ee
The intrinsic scatter $\Delta_{\log M\bh}$ in the mass-dispersion relation is
small enough that it is difficult to distinguish from measurement errors, but
is less than $0.27\dex$ \citep{Tremaine02}. The corresponding correction
factor is therefore between 1 and 1.2, and we shall adopt 1.1. The intrinsic
dispersion in the mass--luminosity relation is $\Delta_{\log M\bh}\simeq
0.5\dex$ (note that only the intrinsic dispersion, not measurement error, is
included in the definition of $\Delta_{\log M\bh}$), so the corresponding
correction is a factor of 2.0. Thus the density estimates (\ref{eq:rhosigmaa})
and (\ref{eq:rhola}) become
\be
\rho\bh(z=0)=(2.5\pm 0.4)\times 10^5\msun\Mpc^{-3},
\label{eq:rhosigma}
\ee
and
\be
\rho\bhL(z=0)=5.8\times 10^5\msun\Mpc^{-3}.
\label{eq:rhol}
\ee
The error estimate in equation (\ref{eq:rhosigma}) represents the combination
of the $\sim$10\% uncertainty in the correction factor 1.1 due to intrinsic
scatter, and the error estimate in equation
(\ref{eq:rhosigmaa}). The error in equation
(\ref{eq:rhol}) is difficult to estimate but is certainly much larger than the
error in equation (\ref{eq:rhosigma}). Earlier
estimates include $\rho\bhL(z=0)=3.5\times 10^5\msun\Mpc^{-3}$
\citep{Salucci99}, $\rho\bhL(z=0)=3.7\times 10^5\msun\Mpc^{-3}$ \citep{MF01b},
and $\rho\bhL(z=0)=(4\pm2)\times 10^5\msun\Mpc^{-3}$ \citep{MS01}.

The estimates (\ref{eq:rhosigma}) and (\ref{eq:rhol}) differ by a factor of
more than two. We believe that equation (\ref{eq:rhosigma}) is substantially
more reliable than equation (\ref{eq:rhol}), since the correlation between
$M\bh$ and $\sigma$ is tighter and the result does not depend on the uncertain
bulge-disk decomposition. Nevertheless, the discrepancy between the two
equations is worth investigating. The problem cannot arise from different
estimates of the local galaxy density, since both results employ the same
galaxy sample
\citep{Bernardi01}; nor can it plausibly arise from errors in the BH mass
determinations, since the mass-dispersion and mass-luminosity relation are
based on the same set of mass estimates (see Tremaine et al. 2002 and
Kormendy \& Gebhardt 2001, respectively). To isolate the problem, it is useful
to examine the Faber-Jackson relation between dispersion and luminosity. The
mean relation derived from a large sample of early-type galaxies
\citep{pru96} by \citet{for99} is
\be
\log(\sigma/\kms)=2.25-0.102(M_B+20),
\label{eq:prug}
\ee
while the mean relation derived from the \citet{Bernardi01} sample is
\be
\log(\sigma/\kms)=2.18-0.100(M_B+20).
\ee
Thus the mean absolute magnitude at a given dispersion in the
\citet{Bernardi01} sample is displaced by $\Delta_1 M_B=-0.7$ from
equation (\ref{eq:prug}). On the other hand, the mean residual in absolute
magnitude from the relation (\ref{eq:prug}) for the 31 early-type galaxies
with BH mass determinations in \citet{Tremaine02} is $\Delta_2 M_B=+0.54$.  A
component of the latter difference arises because equation (\ref{eq:prug})
includes both bulge and disk luminosity in S0 galaxies, while the luminosities
quoted in \citet{Tremaine02} refer only to the bulges in S0
galaxies. Correcting crudely for this using equation (\ref{eq:lfrac}) reduces
$\Delta_2 M_B$ to +0.24. Together, $\Delta_1 M_B$ and $\Delta_2 M_B$ imply that
the mean absolute magnitude at a given dispersion is about 1 magnitude
brighter in the $\sim 10^4$ SDSS galaxies used to determine the
velocity-dispersion and luminosity distribution of nearby galaxies than in the
$\sim 30$ galaxies used to determine the mass-luminosity and mass-dispersion
relation for BHs. This difference is sufficient to explain the difference in
BH density estimates in equations (\ref{eq:rhosigma}) and (\ref{eq:rhol}). 
What is less clear is the origin of this difference. Possible explanations
include selection effects in either the \citet{Bernardi01} sample or the
sample of galaxies with BH masses, or systematic differences in the magnitude
or dispersion measurements between SDSS and pointed galaxy surveys. 

We close this section with comments on the contribution to the estimated BH
density from the faint end of the galaxy luminosity function. The SDSS galaxy
sample is almost complete at the bright end, but the sample is
cut off at the faint end ($M_{g^*}-5\log h\ga -19.0$, cf. Fig.~6 in Bernardi
et al. 2001).  In our calculation we have extrapolated Bernardi et al.'s
distribution of magnitude (or velocity dispersion) to arbitrarily faint
galaxies.  This uncertain extrapolation should not significantly affect our
estimate of BH mass density, as seen from the following arguments.  (i) If we
assume that the distribution of absolute magnitude obtained in the SDSS is
constant for all magnitudes fainter than $\langle M_{g^*}\rangle=-19.65+5\log
h$ (rather than obeying a Gaussian distribution in $M_{g^*}$), the total
luminosity density is increased by less than 15\%.  (ii) The 2dF Galaxy Survey
extends to considerably fainter absolute magnitudes ($M_{\rm b_J}\la -14+5\log
h$) [$M_{\rm b_J}=M_{g_*}+0.16+0.15(g^*-r^*)$ in Norberg et al. 2001] than the
SDSS sample.  The type 1 galaxies in the 2dF Galaxy Survey include all the
early-type galaxies and some spiral galaxies.  The difference between the
total luminosity density obtained by assuming that the distribution of
absolute magnitude of the type 1 galaxies is constant for all magnitudes
fainter than $M_{\rm b_J}=\langle M_{g^*}\rangle+0.16+0.15(g^*-r^*)$ and that
obtained by extrapolating the distribution of absolute magnitude of the type 1
galaxies (cf. Madgwick et al. 2001) is less than 4\%.  In addition, in
\S~\ref{sec:clue} below (see also Figs.~\ref{fig:mass} and \ref{fig:entropy}),
we will focus on comparing the distribution of local high-mass BHs ($\ga
10^8\msun$) with the luminosity function of luminous QSOs, where we need not
consider uncertainties of local BH mass function at the low-mass ends.

%==========================================================================
\subsection{BH mass density accreted during bright QSO phases} \label{subsec:bhacc}
\noindent
Most astronomers believe that QSOs are powered by gas accretion onto BHs, and
in this case the luminosity function of QSOs as a function of redshift
reflects the gas accretion history of local remnant BHs \citep{S82,CT92,SB92}.
We denote the QSO bolometric
luminosity produced by a mass accretion rate $\dot M\acc$ as
$L\bol=\epsilon\dot M\acc c^2=\epsilon\dot M\bh c^2/(1-\epsilon)$, where
$\epsilon$ is the mass-to-energy conversion efficiency (up to 31\% for
thin-disk accretion onto a maximally rotating Kerr BH; Thorne 1974), $\dot
M\bh=\dot M\acc (1-\epsilon)$ is the growth rate of BH mass, and $c$ is the
speed of light.  The comoving BH mass density accreted during bright QSO
phases is given by
\be
\rho\bhaccr\QSO(z)=\int_z^\infty\frac{\d t}{\d z}\d z
\int_0^\infty \frac{(1-\epsilon)L\bol}{\epsilon c^2}\Psi(L,z)\d L,
\label{eq:rhoacc}
\ee
where $\Psi(L,z)$ is the luminosity function of QSOs, defined so that
$\int\Psi(L,z)\d L$ gives the comoving number density of QSOs at redshift $z$
($L$ can be either the bolometric luminosity or the luminosity at a given
band).
In the $B$ band, the bolometric correction $C_B$, defined by
$L\bol=C_B\nu_BL_\nu(B)$, is about 11.8 \citep{Elvis94}.
Here $\nu_BL_\nu(B)$ is the energy radiated at the central
frequency of the $B$ band per unit time and logarithmic interval of frequency.

\begin{figure}
\begin{center}
\includegraphics[width=0.8\textwidth,angle=0]{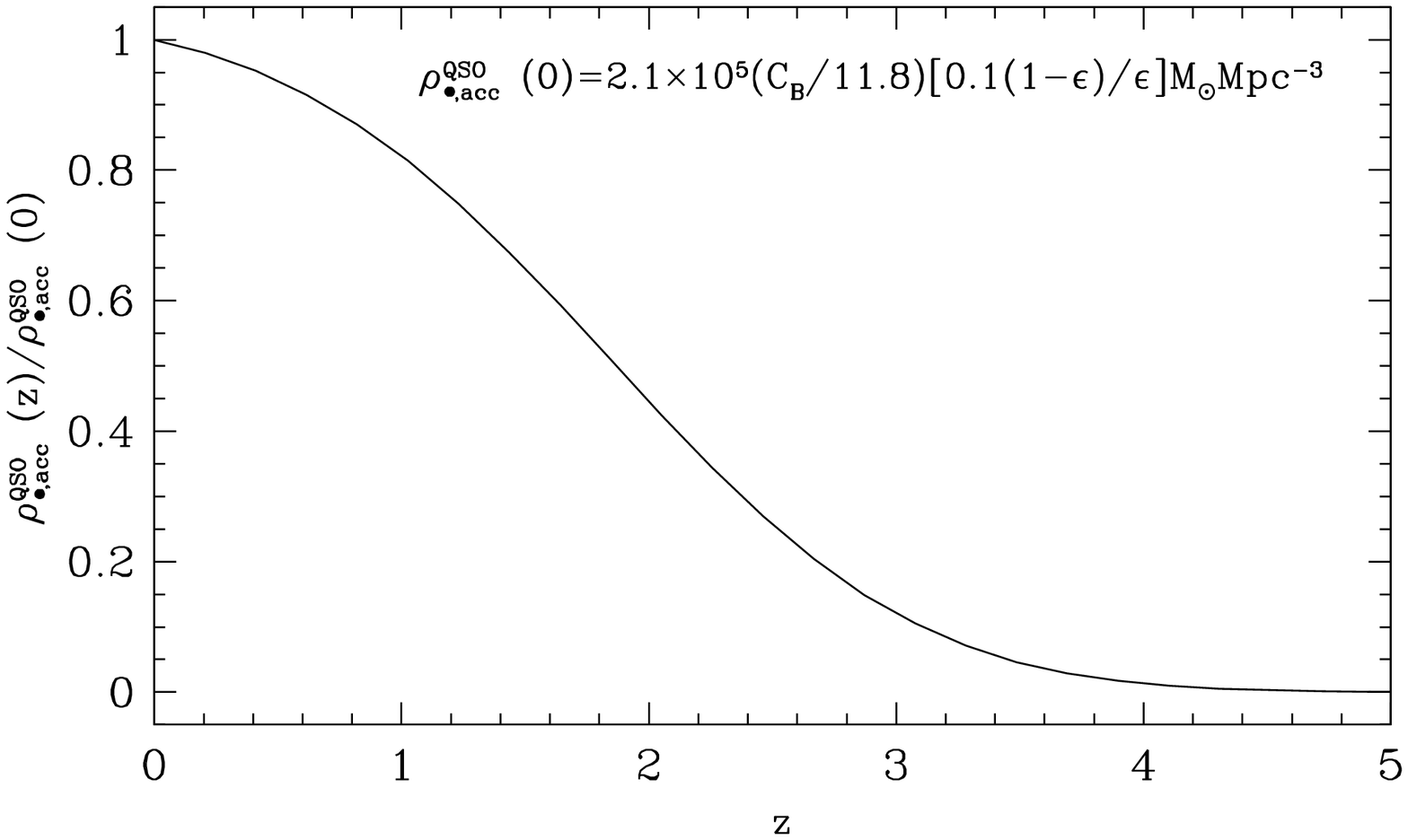}
\caption{The history of the comoving massive BH mass density due to
accretion during optically bright QSO phases.
The QSO luminosity function (eqs.~\ref{eq:QSOLF}--\ref{eq:beta}),
which is only well-determined for $z\la 2.5$,
has been extrapolated to higher redshift.
The mass-to-energy conversion efficiency $\epsilon$ is assumed to
be a constant. See details in \S~\ref{subsec:bhacc}.
}
\label{fig:bhz}
\end{center}
\end{figure}

At $z\la 3$, the QSO luminosity function is often fitted with a double
power law:
\be
\Psi_M(M_B,z)=\frac{\Psi_M^*}
{10^{0.4(\beta_1+1)[M_B-M^*_B(z)]}+10^{0.4(\beta_2+1)[M_B-M^*_B(z)]}},
\label{eq:QSOLF}
\ee
where $\Psi_M(M_B,z)\d M_B$ is the comoving number density of QSOs with
absolute magnitude in the range [$M_B, M_B+\d M_B$] and
$\Psi_M(M_B,z)=\Psi(L_B,z)|\d L_B/\d M_B|=0.92L_B\Psi(L_B,z)$.
\citet{Boyle00} use this functional form to fit the data sets from the
2dF QSO Redshift Survey \citep{Boyle00} and the Large Bright QSO Survey
\citep{HFC95}, which contain over 6000 QSOs.
For example, in our standard model,
$(\Omega_{\rm m},\Omega_\Lambda)=(0.3,0.7)$,
the luminosity function of QSOs with absolute magnitudes $-26<M_B<-23$
and redshifts $0.35<z<2.3$ is described by the following parameters:
\begin{eqnarray}
\Psi_M^*=2.9\times10^{-5}h^3\Mpc^{-3}\mag^{-1},\\
M^*_B(z)=M^*_B(0)-2.5(k_1z+k_2z^2), \label{eq:Mstarz}\\
M^*_B(0)=-21.14+5\log h, k_1=1.36, k_2=-0.27, \label{eq:k1k2}\\
\beta_1=-1.58 {~~\rm and~~}  \beta_2=-3.41. \label{eq:beta}
\label{eq:QSOLFpara}
\end{eqnarray}
The quadratic dependence of the characteristic magnitude $M^*_B(z)$ on $z$
(eqs.~\ref{eq:Mstarz} and \ref{eq:k1k2})
shows an increasing characteristic luminosity with increasing redshift at
low redshift ($z\la 2.5$) and a decline of the characteristic luminosity
at higher redshift, which is suggested by observations
(e.g. Shaver et al. 1996), although the luminosity function is not yet
accurate enough to confirm the decline at $z>2.5$.
The QSO luminosity function over the range $3.6<z<5.0$ provided in
\citet{Fan01} gives a flatter bright-end slope ($\beta_2=-2.5$) than equation
(\ref{eq:QSOLF});
however in our calculations below, we will simply extrapolate equation
(\ref{eq:QSOLF}) to high redshift ($z>3$)
because the detailed QSO luminosity function at $z>3$ does not affect our
final results much
(cf. Figure~\ref{fig:bhz} below or the discussion in Chokshi \& Turner
1992; if we use the luminosity function in Fan et al. 2001, our results
change by less than one percent).

Using equation (\ref{eq:rhoacc}) we may now obtain the accretion history of the
BH mass density arising from optically bright QSO phases, $\rho\bhaccr\QSO(z)$.
The result is shown in Figure~\ref{fig:bhz}, where
the mass-to-energy conversion efficiency $\epsilon$ is assumed constant.
About 56\% of the present density $\rho\bhaccr\QSO(z=0)$ is accreted
when $z<2$ and 90\% when $z<3$, consistent with our claim that
$\rho\bhaccr\QSO(z=0)$ is insensitive to uncertainties in the QSO
luminosity function at high redshift.
The accreted BH mass density during optically bright QSO phases is found to be
\be
\rho\bhaccr\QSO(z=0)=2.1\times 10^5 (C_B/11.8)[0.1(1-\epsilon)/\epsilon]\msun\Mpc^{-3},
\label{eq:rhoacc0}
\ee
which is independent of the Hubble constant and remarkably close to the
value obtained by \citet{CT92} using similar methods,
$\rho\bhaccr\QSO(z=0)=2.2\times 10^5 (C_B/16.5)(0.1/\epsilon)\msun\Mpc^{-3}$.
If the QSOs have a mass-to-energy conversion efficiency $\epsilon\simeq0.1$,
$\rho\bhaccr\QSO(z=0)$ would be close to the local BH
mass density $\rho\bh$ (eq.~\ref{eq:rhosigma}),
which implies that the local BH mass density comes mainly from accretion
during optically bright QSO phases.

The accreted BH mass density can also be inferred from the X-ray and infrared
backgrounds, for example, $\rho^{\rm X}\bh\sim$(6--9)$\times10^5
(0.1/\epsilon)\msun\Mpc^{-3}$ in \citet{FI99},
$\rho^{\rm X}\bh\sim$\- (7.5--16.8)$\times10^5(0.1/\epsilon)\msun\Mpc^{-3}$
in \citet{ERZ02},
$\rho^{\rm IR}_\bullet\sim7.5\times10^5 (0.1/\epsilon)$\- $\msun$$\Mpc^{-3}$
in \citet{HK01}.
Multi-waveband (including optical and far infrared)
observations of {\it Chandra} hard X-ray sources give
$\rho^{\rm HX+OPT+FIR}\bh\sim9\times10^5(0.1/\epsilon)$\- $\msun\Mpc^{-3}$
in \citet{Barger01}.
\citet{FI99} conclude that 85\% of the emitted energy from AGNs
has been absorbed, so that a large fraction of the local BH
mass density must be due to obscured QSOs, which do not contribute to
the estimate in equation (\ref{eq:rhoacc0}).
Thus, their estimate $\rho\bh^{\rm X}$ exceeds $\rho\bhaccr\QSO(z=0)$
by a factor 3--4, and then the comparison of $\rho^{\rm X}\bh$ with
the local BH mass density (eq.~\ref{eq:rhosigma})
implies the efficiency $\epsilon\sim$0.3--0.5, at or beyond the upper limit of
plausible accretion processes, $\epsilon=0.31$ \citep{T74}.
\citet{ERZ02} reached a similar conclusion, $\epsilon\ga 0.15$, by comparing
$\rho^{\rm X}\bh$ with the local BH mass density
[they used an early estimate of the local BH mass
density $3.5\times 10^5\msun\Mpc^{-3}$ from Salucci et al. (1999), a factor
of 1.4 larger than our estimate in eq.~(\ref{eq:rhosigma})].
However, there exist many uncertainties in the estimation of the BH
mass density from the X-ray or infrared background since
the population of AGNs contributing to those backgrounds is not yet clearly
understood, and is not necessarily the same as the optically bright QSOs.

If the local BH mass density comes mainly from optically bright QSOs,
not only the local BH mass density but also the local BH mass distribution
should be consistent with the accretion history determined from QSOs.
The latter comparison is the subject of the next section.

%%%%%%%%%%%%%%%%%%%%%%%%%%%%%%%%%%%%%%%%%%%%%%%%%%%%%%%%%%%%%%%%%%%%%%%%%%%
\section{Some clues from the BH mass distribution} \label{sec:clue}

\noindent
In this section, we employ a continuity equation for the evolution of the
BH distribution to relate the local BH mass function and the QSO
luminosity function.

%==========================================================================
\subsection{Expectations from the continuity equation for BH distributions}
\label{subsec:continu}

\noindent
We define $S$ to be any physical variable associated with a BH (e.g. mass,
entropy etc.) and $n_S(S,t)$ so that $n_S(S,t)\d S$ is the number density
of BHs with $S\rightarrow S+\d S$ at time $t$.
The evolution of the distribution $n_S(S,t)$ is determined by both
BH coalescence (caused by galaxy mergers) and other physical processes
(e.g. gas accretion).
We can use the following continuity equation (cf. Small \& Blandford 1992)
to describe the evolution of the $S$-distribution:
\be
\frac{\partial n_S(S,t)}{\partial t}+
\frac{\partial[n_S(S,t)\langle \dot S \rangle] }{\partial S}=\gamma\merge(S,t),
\label{eq:continu}
\ee
where $\gamma\merge(S,t)$ represents the variation of the BH distribution
caused by mergers, and $\langle \dot S \rangle (S,t)$ is the mean rate of
change of $S$ due to physical processes that conserve the number of BHs.
In equation (\ref{eq:continu}), we ignore the generation of seed BHs, which
are usually assumed to have small masses (e.g. $\la 10^6\msun$).
Since the process of BH mergers is not understood, we shall consider two
extreme cases: one is the classical case,
in which gravitational radiation can be neglected and the total BH mass is
conserved during a merger; and the other is the adiabatic case,
in which gravitational radiation cannot be ignored and the total BH area
(or entropy) is conserved.
We will consider the evolution of the BH mass distribution in the classical
case (i.e., set $S$ in eq.~\ref{eq:continu} to be the
reducible\footnote{The dynamically measured BH mass $M\bh$
(in eqs.~\ref{eq:msigma} and \ref{eq:mlb}) is the reducible mass.
For a BH with reducible mass $M\bh$ and dimensionless spin $a$
($0\le |a|\le 1$), the irreducible mass is given by
$M_{\rm ir}=M\bh(1+\sqrt{1-a^2})^{1/2}/\sqrt{2}$ \citep{MTW73}.
Thus, $M\bh/\sqrt{2}\le M_{\rm ir}\le M\bh$.}
BH mass $M\bh$), and the evolution of the area distribution
in the adiabatic case
(i.e., set $S$ in eq.~\ref{eq:continu} to be the BH area $A\bh$).
In either case, mergers conserve the total value of $S$, so we may consider a time interval $\d t$ in which two BHs
parameterized by $S_1$, $S_2$ merge to form a single BH parameterized by
$S_1+S_2$, i.e.,
\be
\gamma\merge(S',t)~\d t=\delta(S'-S_1-S_2)-\delta(S'-S_1)-\delta(S'-S_2).
\label{eq:s1s2}
\ee
Thus
\be
\int_0^\infty S\gamma\merge(S,t)~\d S=0.
\label{eq:gamma}
\ee

If $\gamma\merge(S,t)$ were known, we could
integrate equation (\ref{eq:continu}) to obtain $n_S(S,t)$, which could be
compared to the local BH mass distribution. Unfortunately the BH merger rate
is very uncertain. We therefore ask a more restricted question: can we find
some function of the mass distribution that can only increase during mergers,
so that even without knowing the merger rate we can establish some
inequalities on the current BH mass distribution and QSO luminosity function?
For example, we would like to choose a function $f(S,S')$ so that
\begin{eqnarray}
\Gamma\merge(S,t)\equiv \int_{0}^\infty f(S,S')\gamma\merge(S',t)~\d S'
\label{eq:Gmerge}
\end{eqnarray}
is positive-definite. By inserting equation (\ref{eq:s1s2}) in equation
(\ref{eq:Gmerge}), we have
\be
\Gamma\merge(S,t)~\d t=f(S,S_1+S_2)-f(S,S_1)-f(S,S_2).
\label{eq:GammaS}
\ee
A sufficient condition that equation (\ref{eq:GammaS}) is positive definite
for $S_1,S_2>0$ is that $f(S,S')/S'$ is
a monotonically increasing function of $S'$ \citep{HLP34}. We shall
choose a specific function $f(S,S')$ below.
By integrating equation (\ref{eq:continu}) over time $t$ and setting the
initial BH $S$-distribution $n_{S}(S,0)=0$, we may
obtain the BH distribution $n_{S}(S,t_0)$ at the present time $t_0$
as follows:
\be
n_{S}(S,t_0)=-\int_0^{t_0}\frac{\partial [n_S(S,t)\langle \dot S \rangle]}
{\partial S}~\d t+\int_0^{t_0} \gamma\merge(S,t)~\d t.
\label{eq:bhnum}
\ee
Now integrate equation (\ref{eq:Gmerge}) from $t=0$ to $t_0$, and use equation
(\ref{eq:bhnum}) to eliminate $\gamma\merge$.
Assuming $f(S,0)=0$, we have
\be
G\local(S,t_0)=G\acc(S,t_0)+\int_0^{t_0}\Gamma\merge(S,t)~\d t,
\label{eq:Gevollocal}
\ee
where
\be
G\local(S,t_0)\equiv\int_0^\infty f(S,S')n_S(S',t_0)~\d S'
\label{eq:GSlocalf}
\ee
is related to the BH $S$-density in local galaxies, and
\be
G\acc(S,t_0)\equiv\int_0^\infty \d S'\int_0^{t_0}
n_S(S',t)\langle \dot S'\rangle \frac{\partial f(S,S')}{\partial S'}~\d t
\label{eq:GSaccf}
\ee
is related to the accretion history of local BHs.
Since $\Gamma\merge(S,t)$ is non-negative, we have
\be
G\local(S,t_0)\ge G\acc(S,t_0).
\label{eq:GSuneq}
\ee
Inequality (\ref{eq:GSuneq}) must hold for every $f(S,S')$ such that
$f(S,S')/S'$ is monotonically increasing in $S'$.
We have assumed that $f(S,0)=0$;
it is also useful for $f(S,S')$ to be continuous, since its derivative
appears in equation (\ref{eq:GSaccf}).
A simple function satisfying these conditions, which we shall use for the
remainder of this paper, is
\be
f(S,S') = \cases{ 0 & for $S'<S$, \cr S'-S & for $S'\ge S$.}
\label{eq:fS}
\ee
In this case,
\be
G\local(S,t_0)=\int_{S}^\infty (S'-S)n_{S}(S',t_0)~\d S',
\label{eq:GSlocal}
\ee
\be
G\acc(S,t_0)=\int_S^\infty \d S'\int_0^{t_0}
n_S(S',t)\langle \dot S' \rangle~\d t.
\label{eq:GSacc}
\ee
With this definition, equation (\ref{eq:gamma}) implies that
$\Gamma\merge(0,t)=0$, so
\be
G\local(0,t_0)=G\acc(0,t_0).
\label{eq:GSeq}
\ee
Using this result we can define a normalized version of the fundamental
inequality (\ref{eq:GSuneq}),
\be
g\local(S,t_0)\equiv\frac{G\local(S,t_0)}{G\local(0,t_0)}\ge\frac{G\acc(S,t_0)}{G\acc(0,t_0)}\equiv g\acc(S,t_0).  \label{eq:gSuneq}
\ee
In the classical case where $S=M_\bullet$,
equation (\ref{eq:GSeq}) is identical to So{\l}tan's argument (1982) relating
the total local energy density in QSOs to the total BH mass density in nearby
galaxies, while the inequality (\ref{eq:gSuneq}) provides a constraint on the
normalized BH mass distribution that is independent of
So{\l}tan's. Inequalities (\ref{eq:GSuneq}) and (\ref{eq:gSuneq}) will be our
principal tools to constrain the accretion history of local BHs.

%--------------------------------------------------------------------------
\subsubsection{Relating the local BH mass function to the QSO luminosity
function} \label{subsubsec:relate}

\noindent
To proceed further, we now assume that the luminosity $L$ during bright QSO
phases is an (increasing) function of only the central BH mass [$\d L(M\bh)/\d
M\bh>0)$], which is plausible since usually we have $L\propto L\bol$ and,
especially for luminous QSOs, $L\bol(M\bh)$ is usually assumed to be near the
Eddington luminosity. In our two extreme cases, we have

\begin{itemize}
\item Classical case: $S\equiv M\bh$.
We have
\be
G\QSO\acc(M\bh,t_0)=\int_{L(M\bh)}^\infty \d L'\int_0^{t_0}
\Psi(L',t)\frac{(1-\epsilon)L\bol'}{\epsilon c^2}~\d t.
\label{eq:GMqso}
\ee
which represents the time-integrated mass density accreted onto BHs with
masses larger than $M\bh$ during optically bright QSO phases.  Equation
(\ref{eq:GMqso}) is independent of the duty cycle of bright QSO phases, so
long as the luminosity during these phases is determined by $M\bh$.  If the
local BH mass density arises entirely by mergers and by accretion during
optically bright QSO phases, then by combining inequality (\ref{eq:gSuneq})
with equation (\ref{eq:GMqso}) we have
\be
G\QSO\acc(M\bh,t_0)=G\acc(M\bh,t_0),
\ee
\be
g\local(M\bh,t_0)\equiv\frac{G\local(M\bh,t_0)}{G\local(0,t_0)}\ge \frac{G\QSO\acc(M\bh,t_0)}{G\QSO\acc(0,t_0)}\equiv g\QSO\acc(M\bh,t_0).
\label{eq:gMuneq}
\ee

\item Adiabatic case: $S\equiv A\bh=16\pi G^2 M_{\rm ir}^2/c^4$,
where $G$ is the gravitational constant and $M_{\rm ir}$ is the irreducible
mass of a BH with reducible mass $M\bh$ (see footnote 1).  We assume that the
dimensionless spin of BHs in QSOs is in an equilibrium state (i.e., $\dot
a=0$), which we denote by $a\qso$. For example, in
the thin-disk accretion model for QSOs
\footnote{We assume that any QSO is due to accretion onto a single
massive BH, rather than accretion onto two or more BHs in the galactic
nucleus, although binary BHs are likely to be present in some
galactic centers (Begelman, Blandford \& Rees 1980; Yu 2002).
If there exists more than one BH in a QSO and each BH is accreting with the
Eddington luminosity, the mass $M\bh$ obtained from the total QSO luminosity
will represent the total mass of the BHs, and our arguments in the classical
case are unaffected.
However, in the adiabatic case, the entropy or area increase
associated with a given mass accretion rate depends on the number of BHs.},
the dimensionless spin $a$ increases from 0 to reach the equilibrium
value $a\qso=0.998$ after the BH accretes $\sim 1.5$ times its initial mass
(Thorne 1974).
Thus, in the equilibrium spin state,
we have the area (or entropy) variation rate of a BH as follows:
\be
\dot A\bh=32\pi G^2M_{\rm ir}\dot M_{\rm ir}/c^4=
16\pi G^2M\bh\dot M\bh\big(1+\sqrt{1-a\qso^2}\big)/c^4 \qquad (\dot a=0).
\label{eq:dotA}
\ee
If the local BH entropy density arises entirely by mergers and by accretion
during optically bright QSO phases, then equation (\ref{eq:GSacc}) can be
written as
\be
G\acc(A\bh,t_0)=\int_{{\cal L}(M\ir)}^\infty \d L'
\int_0^{t_0}\Psi(L',t)\dot A\bh(M\bh',a\qso)~\d t,
\label{eq:GAqso}
\ee
where the function ${\cal L}(M\ir)$ is defined by
${\cal L}(M\ir)\equiv L(M\bh)$.
It proves useful to use the irreducible mass rather than the area as the
independent variable, so we define
\be
{\cal G}\AQSO\acc(M\ir,t_0)\equiv G\acc(A\bh,t_0),\qquad\hbox{where}\qquad  
 A\bh=16\pi G^2 M_{\rm ir}^2/c^4.
\label{eq:GAqsoo}
\ee

In the adiabatic case, equation (\ref{eq:GSlocal}) may be written as
\be
G\local(A\bh,t_0)=\int_{M\bh(A\bh,a)}^\infty [A\bh(M\bh',a)-A\bh]
n_{M\bh}(M\bh',t_0)\d M\bh'
\label{eq:GAlocal}
\ee
where for the moment we assume that all the local BHs have the same spin $a$.
Unfortunately, the BH spin parameter $a$ in dead quasars is unknown (in
particular, there is no reason to suppose that it equals the equilibrium spin
parameter $a\qso$ for accreting BHs). However, for a given value of the
reducible mass $M\bh$, both the irreducible mass $M\ir$ and the area $A\bh$
decrease with increasing $a$. Thus we have the inequality 
\be
G\local(A\bh,t_0)\le\int_{M\bh(A\bh,a)}^\infty [A\bh(M\bh',a=0)-A\bh]
n_{M\bh}(M\bh',t_0)\d M\bh'.
\label{eq:GAlocala}
\ee
Once again introducing the irreducible mass rather than the area as the
independent variable, we define
\be
{\cal G}^{A\bh}\local(M\ir,t_0)\equiv\int_{M\ir}^\infty
[A\bh(M\bh',a=0)-A\bh(M\ir)]n_{M\bh}(M\bh',t_0)\d M\bh'.
\label{eq:GAlocalb}
\ee
The integral on the right side of equation (\ref{eq:GAlocala}) differs from
the integral on the right side of equation (\ref{eq:GAlocalb}) only in the
lower limit of the integrand. Since $M\ir\le M\bh$ (see footnote 1), the
second integral has a larger integration range and therefore is larger. Thus
\be
G\local(A\bh,t_0)\le {\cal G}^{A\bh}\local(M\ir,t_0).
\label{eq:GAlocalc}
\ee
The inequality (\ref{eq:GAlocalc}) is an equality if and only if BHs in local
galaxies are all Schwarzschild BHs ($a=0$).
By using inequalities (\ref{eq:GSuneq}) and (\ref{eq:GAlocalc}) and equations
(\ref{eq:GAqso}) and (\ref{eq:GAqsoo}), we have finally:
\be
{\cal G}^{A\bh}\local(M\ir,t_0)\ge {\cal G}\AQSO\acc(M\ir,t_0).
\label{eq:GAuneq}
\ee
Note that inequality (\ref{eq:GAuneq}) still holds even if we discard the
assumption made after equation (\ref{eq:GAlocal}) that all local BHs have the
same spin parameter. 

\end{itemize}

To estimate $G\local$ and ${\cal G}\local^{A\bh}$,
we proceed as follows.
High-luminosity galaxies are mainly early-type galaxies
(see, for example, Figure~4.14 in Binney \& Merrifield 1998
or Figure~7 in Madgwick et al. 2001);
moreover, the mass of the central BH is approximately proportional to the
luminosity of the hot stellar component
(cf. eq.~\ref{eq:mlb}), which is less than $\sim$30\% of the total
luminosity for spiral galaxies \citep{FHP98};
hence high-mass BHs ($M\bh\ga 10^8\msun$, corresponding to
$M_{B,\rm hot}<-20$) are mostly in early-type galaxies.
The local BH mass distribution in early-type galaxies can be obtained
from their velocity dispersions and the mass--dispersion correlation.
By analogy with equations (\ref{eq:GSlocal}) and (\ref{eq:GAlocal}),
we define
\be
G\early\local(M\bh,t_0)\equiv\int_{M\bh}^\infty (M\bh'-M\bh)n\early_{M\bh}(M\bh',t_0)\d M\bh'
\label{eq:GMearly}
\ee
and
\be
{\cal G}\local\Aearly(M\ir,t_0)\equiv \int_{M\bh}^\infty
[A\bh(M\bh',a=0)-A\bh(M\ir)]n_{M\bh}\early(M\bh',t_0)~\d M\bh'.
\label{eq:GAearly}
\ee
For high-mass BHs ($M\bh\ga 10^8\msun$), we have
$G\early\local(M\bh,t_0)\simeq G\local(M\bh,t_0)$ and
${\cal G}\Aearly\local(M\bh,t_0)$ \- $\simeq{\cal G}\local^{A\bh}(M\bh,t_0)$.
In addition, as seen from inequalities (\ref{eq:GSuneq}), (\ref{eq:gMuneq})
and (\ref{eq:GAuneq}), we have
\be
G\early\local(M\bh,t_0)\ga G\QSO\acc(M\bh,t_0),
\label{eq:GMearlyuneq}
\ee
\be
g\local\early(M\bh,t_0)\equiv\frac{G\early\local(M\bh,t_0)}{G\local(0,t_0)}\ga
\frac{G\QSO\acc(M\bh,t_0)}{G\QSO\acc(0,t_0)}\equiv g\acc\QSO(M\bh,t_0)
\qquad (M\bh\ga 10^8\msun),
\label{eq:gMearlyuneq}
\ee
in the classical case
(with approximate equality if and only if mergers are neglected)
and
\be
{\cal G}\Aearly\local(M\ir,t_0)\ga {\cal G}\acc\AQSO(M\ir,t_0) \qquad (M\ir\ga 10^8\msun)
\label{eq:GAearlyuneq}
\ee
in the adiabatic case (with approximate equality if and only if mergers
are neglected and BHs in local galaxies are all Schwarzschild BHs).
The approximate nature of the inequalities arises solely from our restriction
to early-type galaxies.

%==========================================================================
\subsection{Comparison with observations}
\label{subsec:comparison}

\noindent
We now compare inequalities (\ref{eq:GMearlyuneq})--(\ref{eq:GAearlyuneq})
with the observations. Throughout this section, all the variables related to
the distribution of BH mass and entropy density, $G\early\local(M\bh,t_0)$,
$g\local\early(M\bh,t_0)$ and ${\cal G}\Aearly\local(M\ir,t_0)$, are obtained
by using the mass--dispersion relation (eq.~\ref{eq:msigma}) and ignoring the
small intrinsic scatter in this relation.

%--------------------------------------------------------------------------
\subsubsection{The classical case} \label{subsubsec:classical}
\begin{figure}
\begin{center}
\includegraphics[width=0.8\textwidth,angle=0]{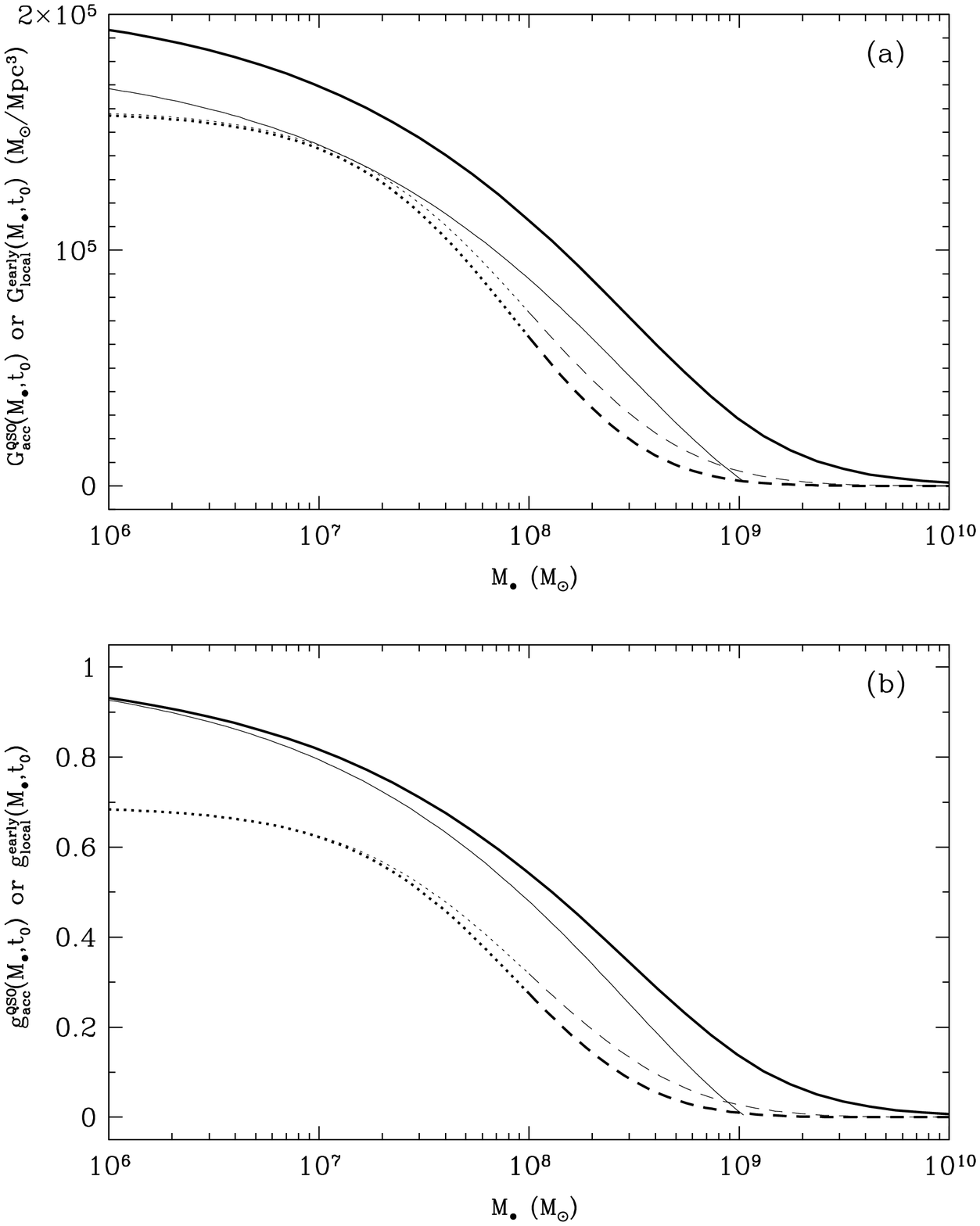}
\caption{Panel (a):
Variables related to the BH mass density as a function of BH mass $M\bh$.  The
thick solid line shows $G\QSO\acc(M\bh,t_0)$ contributed by QSOs
(eq.~\ref{eq:GMqso}), where QSOs are assumed to have the Eddington luminosity
and the mass-to-energy efficiency $\epsilon$
is given by $\epsilon/(1-\epsilon)=0.1$.  The dashed line ($M\bh\ga
10^8\msun$) and the dotted line ($M\bh\la 10^8\msun$) show
$G\local\early(M\bh,t_0)$ obtained from the SDSS sample of nearby early-type
galaxies (eq.~\ref{eq:GMearly}).  In the dotted region $G\early\local$
underestimates $G\local$ because of the contribution from BHs in spiral
bulges.  The thick dotted and dashed lines show the result obtained by using
the mass--dispersion relation in \citet{Tremaine02}, and the thin lines show
the result obtained by using the relation in \citet{MF01a}.  Panel (b):
Variables related to the normalized BH mass density as a function of BH mass
$M\bh$, i.e., $g\QSO\acc(M\bh,t_0)$ and $g\local\early(M\bh,t_0)$
(eq.~\ref{eq:gMearlyuneq}).  QSOs are assumed to have the Eddington luminosity
and the curves are independent of the
mass-to-energy efficiency $\epsilon$, so long as it is a constant.  The line
types have the same meanings as those in Panel (a).  At the low-mass end,
$g\early\local(M\bh,t_0)$ approaches $(f_{\rm E}+f_{\rm bulge,S0})/f_{\rm
hot}\simeq 0.69$ (see \S~\ref{subsec:bhdenlocal}).  For high BH masses
($M\bh\ga 10^8\msun$), inequalities (\ref{eq:GMearlyuneq}) and
(\ref{eq:gMearlyuneq}) imply that the thick solid line should be equal to the
dashed one in panels (a) and (b) if mergers are unimportant, and lower if
mergers are important, which is inconsistent with the observations.  The thin
solid line represents an estimate of $G\QSO\acc(M\bh,t_0)$ in panel (a) [or
$g\QSO\acc(M\bh,t_0)$ in panel (b)] obtained by cutting off the QSO luminosity
function at $L\bol\ga L\Edd(10^9\msun)$, and this estimate is still
significantly higher than the dashed lines and inconsistent with inequality
(\ref{eq:GMearlyuneq}) (or inequality \ref{eq:gMearlyuneq}).  See discussion
in \S~\ref{subsubsec:classical}.  }
\label{fig:mass}
\end{center}
\end{figure}

\noindent
In the classical case, the total BH mass is conserved during mergers,
which is implicitly assumed in So{\l}tan's argument and
many subsequent discussions of the growth of BHs.
In \S~\ref{sec:bhden}, we have obtained that
the total BH mass density in local galaxies is consistent with that
accreted during optically bright QSO phases,
if the mass-to-energy efficiency $\epsilon\simeq 0.1$,
i.e., $G\local(0,t_0)\simeq G\QSO\acc(0,t_0)$
(cf. eqs.~\ref{eq:rhosigma} and \ref{eq:rhoacc0}).
In contrast, inequality (\ref{eq:gMearlyuneq}) involves normalized quantities
$g(M\bh,t_0)=G(M\bh,t_0)/G(0,t_0)$ and therefore provides a constraint
independent of So{\l}tan's argument.
[For example, note that the normalized quantity $g\acc(M\bh,t_0)$ in equation
(\ref{eq:gMearlyuneq}) is independent of the efficiency $\epsilon$ if
$\epsilon$ is a constant.]

Figure~\ref{fig:mass}(a) shows $G\early\local(M\bh,t_0)$ derived from the
SDSS sample of early-type galaxies (dotted and dashed lines,
eq.~\ref{eq:GMearly}) and $G\QSO\acc(M\bh,t_0)$
derived from optically bright QSOs (thick solid line, eq.~\ref{eq:GMqso}).
The QSOs are assumed to have the Eddington luminosity and
the efficiency $\epsilon$ is given by $\epsilon/(1-\epsilon)=0.1$.
Figure~\ref{fig:mass}(b) shows the normalized quantity
$g\early\local(M\bh,t_0)$ derived from the
SDSS sample of early-type galaxies (dotted and dashed lines,
eq.~\ref{eq:gMearlyuneq}) and $g\QSO\acc(M\bh,t_0)$ derived from optically
bright QSOs (thick solid line, eq.~\ref{eq:gMearlyuneq})
using the same assumptions.
The dotted line in panel (a) [or panel (b)] is used for
$G\local\early(M\bh,t_0)$ [or $g\local\early(M\bh,t_0)$] in the region
$M\bh\la 10^8\msun$ where spiral bulges can make a substantial contribution,
so that $G\local\early$ (or $g\local\early$) is only a lower bound to
$G\local$ (or $g\local$).
For $G\early\local(M\bh,t_0)$ and $g\early\local(M\bh,t_0)$,
we show separately the results obtained by using the mass--dispersion
relations in \citet{Tremaine02} (thick dotted and dashed line)
and \citet{MF01a} (thin dotted and dashed line).
Figure~\ref{fig:mass} shows that for high BH masses ($M\bh\ga10^8\msun$),
$G\QSO\acc(M\bh,t_0)$ [or $g\QSO\acc(M\bh,t_0)$] (thick solid line) is
larger than $G\early\local(M\bh,t_0)$ [or $g\early\local(M\bh,t_0)$]
(dashed lines) by a factor of more than two, which is inconsistent with
inequalities (\ref{eq:GMearlyuneq}) (or \ref{eq:GSuneq}) and
(\ref{eq:gMearlyuneq}).

This inconsistency is robust to a number of changes in our assumptions.
For example:
\begin{itemize}
\item We have assumed that QSOs radiate at the Eddington limit;
if they have sub-Eddington luminosity (e.g. one tenth of the Eddington
luminosity in Ciotti, Haiman \& Ostriker 2001), the solid lines in
Figures~\ref{fig:mass}(a) and (b) would shift to higher mass
and the discrepancy would become worse.

\item The QSO luminosity function is uncertain at the bright end, and the
mass--dispersion relation is uncertain at the high-mass end.
However, even if we cut off the QSO luminosity function at
$L\bol\ga L\Edd(10^9\msun)\simeq 10^{47}{~\rm erg~s^{-1}}$
($L\Edd$: Eddington luminosity),
the new $G\QSO\acc(M\bh,t_0)$ [or $g\QSO\acc(M\bh,t_0)$]
(thin solid line in Figure~\ref{fig:mass})
is still significantly higher than $G\early\local(M\bh,t_0)$
[or $g\early\local(M\bh,t_0)$] for $10^8\msun\la M\bh\la 10^9\msun$.

\item If there exist mechanisms other than accretion during bright
QSO phases [e.g. accretion via advection dominated accretion flow (ADAF),
see discussion in Haehnelt, Natarajan \& Rees (1998); or accretion of
non-baryonic dark matter] that contribute to the local BH mass density,
then $G\QSO\acc(M\bh,t_0)$ is only a lower limit to $G\acc(M\bh,t)$,
and the inconsistency with inequality (\ref{eq:GSuneq}) for high-mass BHs
would become worse.

\item Study of the X-ray background suggests that a majority of the
total BH mass density is contributed by obscured accretion \citep{FI99}.
If the number ratio of obscured QSOs to optically bright QSOs is independent
of BH mass (e.g., in the standard unification model for AGNs,
obscuration is a purely geometrical effect
and obscured fraction has no relation to central BH mass),
$g\acc(M\bh,t_0)$ would be unchanged, and the inconsistency in
Figure~\ref{fig:mass}(b) would remain.

\end{itemize}

One possible resolution to the inconsistency shown in
Figure~\ref{fig:mass} is that the mass-to-energy conversion efficiency
$\epsilon$ depends on QSO luminosity. In particular, the average efficiency in
luminous QSOs (e.g. $L\bol\ga 10^{46}\ergs$, which corresponds to
the Eddington luminosities of BHs with $M\bh\ga 10^8\msun$) must be $\ga 0.2$
to make the BH mass density accreted during optically bright QSO phases as
large as that in local galaxies (cf. Fig.~\ref{fig:mass}a).  Note that the
required average efficiency is close to the maximum efficiency ($\sim 0.3$ in
Thorne 1974) allowed in thin-disk accretion models. This result suggests that
other accretion mechanisms are less significant than accretion during
optically bright QSO phases, and we may expect that obscured accretion
\citep{FI99} suggested by the X-ray background is not important for the growth
of high-mass BHs, which is consistent with the scarcity of Type II QSOs.

If the growth of low-mass BHs also occurs mainly during optically bright QSO
phases, then less luminous QSOs ($L\bol\la 10^{46}\ergs$) must
accrete with an efficiency less than $\sim 0.1$ to maintain the consistency
between the local BH mass density and the BH mass density accreted during
bright QSO phases (cf. eq.~\ref{eq:GSeq}).  If on the other hand, obscured
accretion contributes significantly to the total BH mass density \citep{FI99},
less luminous QSOs may have efficiency $\sim 0.1$ or even greater.  In this
case, the obscured fraction must be larger for low-mass BHs than for high-mass
BHs, which might suggest that obscured accretion occurs mainly for small BHs
in young growing galaxies as described in the model by \citet{F99}.

Other possible resolutions to the inconsistency shown in Figure \ref{fig:mass}
include:

\begin{itemize}

\item Not all massive BHs may reside in galactic centers; for at least three
reasons: (i) BHs may be left in the halo after galaxy mergers; however,
\citet{Y02} shows that the orbital decay time from dynamical friction for BHs
in the galactic halo is generally much less than a Hubble time, so long as the
orbiting BHs remain part of the tidally stripped remnant of their
original host galaxy. (ii) Incomplete merging of binary BHs formed in mergers
is expected to result in binary BHs with semimajor axes in the range
$10^{-3}$--$10\pc$ depending on the velocity dispersion and shape of the host
galaxy, and the masses of the BHs \citep{Y02}; however, in most cases the
separation is so small that the binary should look like a single merged BH at
current telescope resolutions. (iii) BHs can be ejected from galaxy centers
through either three-body interactions with other massive BHs (e.g. Valtonen
1996) or gravitational radiation reaction during BH coalescence (e.g. Rees
2001).

\item The bolometric correction that we have used, $C_B=11.8$, may be too
large; however, the uncertainty in this estimate of the bolometric correction
\citep{Elvis94} is only $\pm4.3$, and the smallest bolometric correction in
the Elvis et al.\ sample is 5.5, so it is unlikely that errors in the
bolometric correction can lower the required efficiency by more than a factor
of two.

\item Luminous QSOs may accrete with super-Eddington luminosities 
\citep{B01,B02}.

\end{itemize}

%--------------------------------------------------------------------------
\subsubsection{The adiabatic case} \label{subsubsec:entropy}

\begin{figure}
\begin{center}
\includegraphics[width=0.8\textwidth,angle=0]{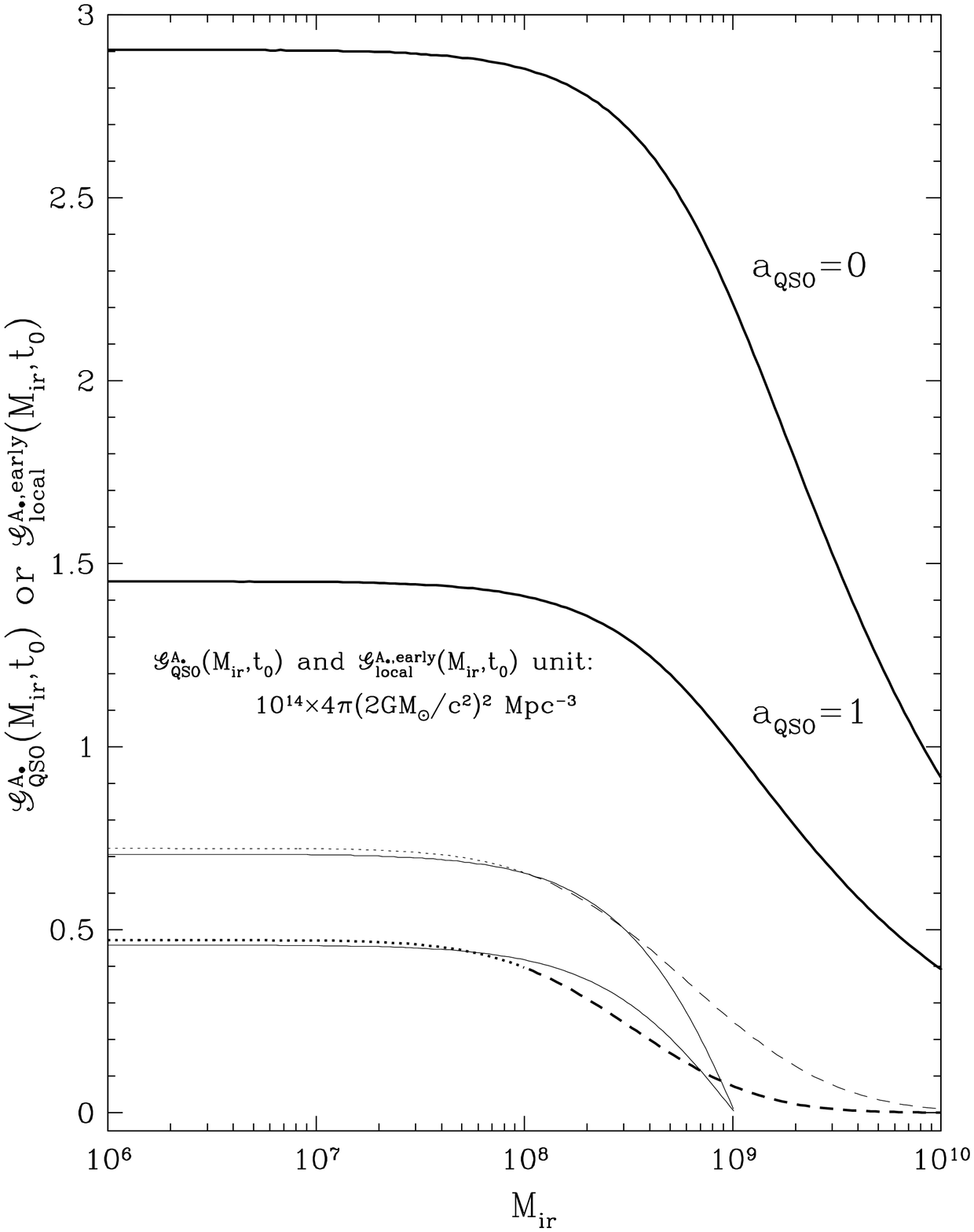}
\caption{
Variables related to the BH entropy density as a function of BH irreducible
mass $M\ir$,
i.e., ${\cal G}\AQSO\acc(M\ir,t_0)$ (eq.~\ref{eq:GAqsoo}, thick solid lines)
and ${\cal G}\Aearly\local(M\ir,t_0)$ (eq.~\ref{eq:GAearly}, dotted and dashed
lines). QSOs are assumed to have the Eddington luminosity and an efficiency
$\epsilon$ given by $\epsilon/(1-\epsilon)=0.1$.  The dotted/dashed lines and
the thin/thick lines have similar meanings to those in Figure~\ref{fig:mass}.
The spin of BHs in QSOs is assumed to be in an equilibrium state $a\qso$.  The
upper thin/thick solid lines are obtained by assuming that all the BHs in QSOs
are Schwarzschild BHs ($a\qso=0$) and the lower thin/thick solid lines assume
that the BHs are maximally rotating ($a\qso=1$).  For high BH masses ($M\ir\ga
10^8\msun$), inequality (\ref{eq:GAearlyuneq}) implies that the thick solid
line should not be higher than the dashed one.
After cutting off the QSO luminosity function at $L\bol\ga
L\Edd(10^9\msun)$, the upper thin solid line for $a\qso=0$ is still significantly
higher than the thick dashed line, which
implies that the BHs in QSOs must be rapidly rotating.  See discussion in
\S~\ref{subsubsec:entropy}.  }
\label{fig:entropy}
\end{center}
\end{figure}

\noindent
BH mass may not be conserved during BH mergers, since mass can be radiated as
gravitational waves. According to the second law of BH thermodynamics, BH
mergers never decrease total BH entropy.  Here, we study an extreme case in
which the BH entropy is conserved during BH mergers.

Figure~\ref{fig:entropy} shows the variables related to the BH
entropy density as a function of BH irreducible mass that we introduced in
\S~\ref{subsec:continu}: ${\cal G}\Aearly\local(M\ir,t_0)$
obtained from the SDSS sample of early-type galaxies (dotted and dashed lines,
see eq.~\ref{eq:GAearly}) and ${\cal G}\AQSO\acc(M\ir,t_0)$ obtained from
optically bright QSOs (thick solid lines, see eq.~\ref{eq:GAqsoo}).  As usual
QSOs are assumed to have the Eddington luminosity and $\epsilon$ is given
by $\epsilon/(1-\epsilon)=0.1$
(eqs.~\ref{eq:QSOLF}--\ref{eq:QSOLFpara}).  The spin of BHs in QSOs is assumed
to be in an equilibrium state, $\dot a=0$,
$a\equiv a\qso$. The upper and lower thick solid lines represent the results
obtained by assuming that all the BHs in QSOs are Schwarzschild
($a\qso=0$) and maximally rotating Kerr BHs ($a\qso=1$),
respectively.  The dotted lines are used for ${\cal G}\Aearly\local(M\ir,t_0)$
in the region $M\ir\la 10^8\msun$ where spiral bulges can make a substantial
contribution, so that ${\cal G}\Aearly\local$ is only a lower bound to ${\cal
G}^{A\bh}\local$.  For ${\cal G}\Aearly\local(M\ir,t_0)$, we show separately
the results obtained by using the mass--dispersion relations in
\citet{Tremaine02} (thick dotted and dashed line) and in \citet{MF01a} (thin
dotted and dashed line).  The functions ${\cal G}\Aearly\local(M\ir,t_0)$
obtained from these two relations differ by a factor of two, much more than
the analogous curves in Figure~\ref{fig:mass}.  The difference is mainly
caused by BHs with $M\bh\ga 10^9\msun$ because the BH area (or entropy)
$A\bh\propto M\ir^2$ so the results are quite sensitive to the
mass--dispersion relation at the high-mass end.  Figure~\ref{fig:entropy}
shows that the thick solid lines [i.e., ${\cal G}\AQSO\acc(M\ir,t_0)$] are
much higher than the dashed lines [i.e., ${\cal G}\Aearly\local(M\ir,t_0)$],
no matter whether we choose $a\qso=0$ or $a\qso=1$ and for both mass--dispersion relations,
which is inconsistent with inequality (\ref{eq:GAearlyuneq}). This
inconsistency may be partly caused by uncertainties in the QSO luminosity
function at the bright end.  To investigate the effects of uncertainties in
the QSO luminosity function, we cut off the luminosity function at $L\bol\ga
L\Edd(10^9\msun)$ (thin solid lines in Figure~\ref{fig:entropy}).  For
consistency, we must then use the Tremaine et al. mass--dispersion
relation since it gives ${\cal G}\Aearly\local(10^9\msun,t_0)\simeq0$.  We see
from Figure~\ref{fig:entropy} that the new ${\cal G}\acc\AQSO(M\ir,t_0)$ for
$a\qso=0$ (upper thin solid line) is still significantly higher than the thick
dashed line, which is inconsistent with inequality (\ref{eq:GAearlyuneq}).
Considering that the maximum efficiency for thin-disk accretion onto
Schwarzschild BHs is less than 0.1, a more realistic ${\cal
G}\acc\AQSO(M\ir,t_0)$ for $a\qso=0$ (i.e., $\epsilon\sim 0.06$) should be even
higher than that shown in Figure~\ref{fig:entropy} and the discrepancy between
the upper thin solid line and the thick dashed line would become worse.
However, the new ${\cal G}\acc\AQSO(M\ir,t_0)$ for $a\qso=1$ (lower thin solid
line) is close to or only slightly higher than the thick dashed line, which is
not seriously inconsistent with inequality (\ref{eq:GAearlyuneq}), and any
inconsistency can be reduced further because rapidly rotating BHs can have
efficiencies substantially larger than 0.1.  Thus, Figure~\ref{fig:entropy}
suggests that: it is impossible that all BHs in luminous QSOs ($L\bol\ga
10^{46}\ergs$) are Schwarzschild BHs; BHs in most luminous QSOs
are rapidly rotating Kerr BHs; and luminous QSOs are accreting at an
efficiency $\ga 0.1$ (cf. Elvis, Risaliti \& Zamorani 2002).

Possible solutions to the inconsistency (mainly for the comparison of
local BHs with QSOs having $a\qso=0$) shown in Figure~\ref{fig:entropy} have been
discussed in \S~\ref{subsubsec:classical}. An additional possibility is that
there exists more than one BH accreting materials in a QSO.  In this case, the
entropy or area increase associated with a given mass accretion rate depends
on the number of BHs (see footnote 2) and equation (\ref{eq:dotA}) is an
overestimate of the area increase.

In the adiabatic case, we do not discuss the roles of obscured accretion
(or other accretion mechanisms) on the BH entropy density
since the total BH entropy obtained from the X-ray background is not clear
and also the conclusion may be affected by the uncertainty of the
mass--dispersion relations at the high BH-mass end or the QSO luminosity
function at the bright end.
%==========================================================================
\subsection{Mass-to-energy conversion efficiency and mean lifetime of QSOs}
\label{subsec:efftime}

\noindent
In \S~\ref{subsec:comparison}, we have argued that the growth of BHs with
$M\bh>10^8\msun$ mainly occurs during optically bright QSO phases
and the mass-to-energy conversion efficiency of luminous QSOs should be
larger than 0.1.
In this subsection, we will ignore BH mergers [the term $\gamma\merge(M\bh,t)$
in eq.~\ref{eq:continu}] and estimate in more detail the mass-dependent
efficiency of QSOs that is required to match the local BH mass function.
We will also give an estimate of the mean lifetime of QSOs.

%--------------------------------------------------------------------------
\subsubsection{Mass-to-energy conversion efficiency} \label{subsubsec:eff}
\begin{figure}
\begin{center}
\includegraphics[width=0.75\textwidth,angle=0]{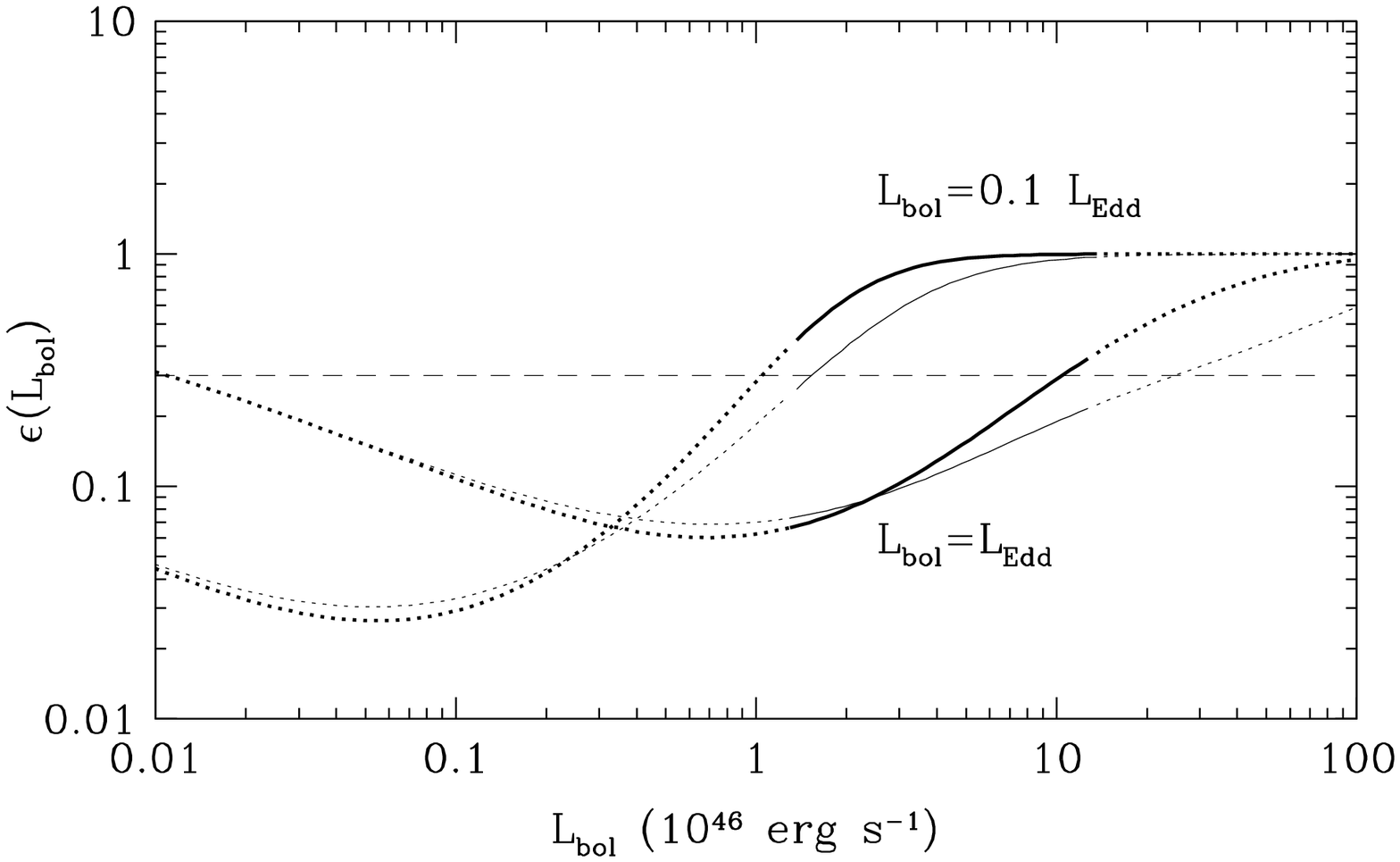}
\caption{The (bolometric) luminosity-dependent efficiency of QSOs
$\epsilon(L\bol)$ if mergers are unimportant (eq.~\ref{eq:eff}). 
The solid lines in the range $10^{46}\ergs\la L\bol\la 10^{47}\ergs$
[corresponding to the BH mass in the range
$10^8(L\Edd/L\bol)\msun \la M\bh\la 10^9(L\Edd/L\bol)\msun$]
show the region in which the observational constraints are
fairly secure, and the dotted lines outside this region show the results
obtained by extrapolating the BH mass function in local galaxies and the QSO
luminosity function. The thick dotted and solid lines are obtained by using
the mass-dispersion relation in \citet{Tremaine02}, and the thin dotted and
solid lines are obtained by using the relation in \citet{MF01a}.
The lower thick/thin lines represent the results by assuming $L\bol=L\Edd$,
and the upper ones represent the results by assuming $L\bol=0.1L\Edd$.
If $L\bol=0.1L\Edd$, the efficiency is required to be higher than 0.3
(the maximum efficiency for thin-disk accretion onto BHs) or even approach 1.
See \S~\ref{subsubsec:eff} for discussion. 
}
\label{fig:eff}
\end{center}
\end{figure}

\noindent
We assume that the mass-to-energy conversion efficiency is a function of BH
mass $\epsilon(M\bh)$, or a function of QSO bolometric luminosity
$\epsilon(L\bol)$ since $L\bol$ is assumed to be a function of only
BH mass $M\bh$. By replacing the variable $S$ in equation (\ref{eq:continu})
by $M\bh'$ and ignoring the term $\gamma\merge$, we have
\be
\frac{\partial n_{M\bh}(M\bh',t)}{\partial t}+
\frac{\partial }{\partial M\bh'}
\left[\Psi(L,t)\frac{\d L}{\d M\bh'}\frac{(1-\epsilon)L\bol(M\bh')}{\epsilon c^2}\right]=0.
\label{eq:continuM}
\ee
By integrating equation (\ref{eq:continuM}) over $M\bh'$ from $M\bh$ to
$\infty$ and over $t$ from 0 to $t_0$, we find
\be
\epsilon(L\bol)=\eta(L\bol)/[1+\eta(L\bol)],
\label{eq:eff}
\ee
where
\be
\eta(L\bol)\equiv \left(\frac{\d L}{\d M\bh}\right)\frac{L\bol}{c^2}
\frac{\int_0^{t_0} \Psi(L,t)~\d t} {\int_{M\bh(L\bol)}^\infty n(M\bh',t_0)\d M\bh'}.
\ee
By inserting the QSO luminosity function and the BH
mass function in local early-type galaxies in equation (\ref{eq:eff}),
and assuming that QSOs have a fixed fraction of the Eddington luminosity (we
examine two cases, 0.1 and 1), we obtain the
efficiency $\epsilon(L\bol)$ in Figure~\ref{fig:eff}.
The solid lines in the interval $10^{46}\ergs\la L\bol\la 10^{47}\ergs$
[corresponding to the BH mass in the range
$10^8(L\Edd/L\bol)\msun\la M\bh\la 10^9(L\Edd/L\bol)\msun$]
denote the region where the observational constraints are fairly strong.
The dotted lines at the luminous and faint ends
are obtained by extrapolating the BH distribution in early-type galaxies
or the QSO luminosity function.
We also show separately the results obtained by using the mass--dispersion
relations in \citet{Tremaine02} (thick dotted and solid lines)
and in \citet{MF01a} (thin dotted and solid lines).
The lower thick/thin solid and dotted lines represent the results by assuming 
$L\bol=L\Edd$; and the upper ones represent the results
by assuming $L\bol=0.1L\Edd$. 
The fact that the efficiency is required to be higher than 0.3
(the maximum efficiency for thin-disk accretion onto BHs) or even approach 1
when $L\bol=0.1L\Edd$ suggests that the assumption of
sub-Eddington luminosity for luminous QSOs is unrealistic, a conclusion
derived by different arguments in \S~\ref{subsubsec:classical}.
Figure~\ref{fig:eff} shows that even for
$L\bol=L\Edd$, when $L\bol$ approaches $10^{47}\ergs$ 
the mass-to-energy efficiency may be slightly larger than 0.3
(the maximum efficiency for thin-disk accretion onto BHs).
As described in \S~\ref{subsubsec:classical},
this unrealistic $\epsilon$ may be caused by
(i) ignoring BH mergers,
(ii) ignoring ejection of BHs from galactic nuclei or BHs left in the halo
after a galaxy merger,
(iii) uncertainties in the mass--dispersion relation at the high-mass
end or the QSO luminosity function at the bright end, or
(iv) ignoring the possibility of super-Eddington luminosities.

%==========================================================================
\subsubsection{Mean lifetime} \label{subsubsec:lifetime}

\begin{figure}
\begin{center}
\includegraphics[width=0.75\textwidth,angle=0]{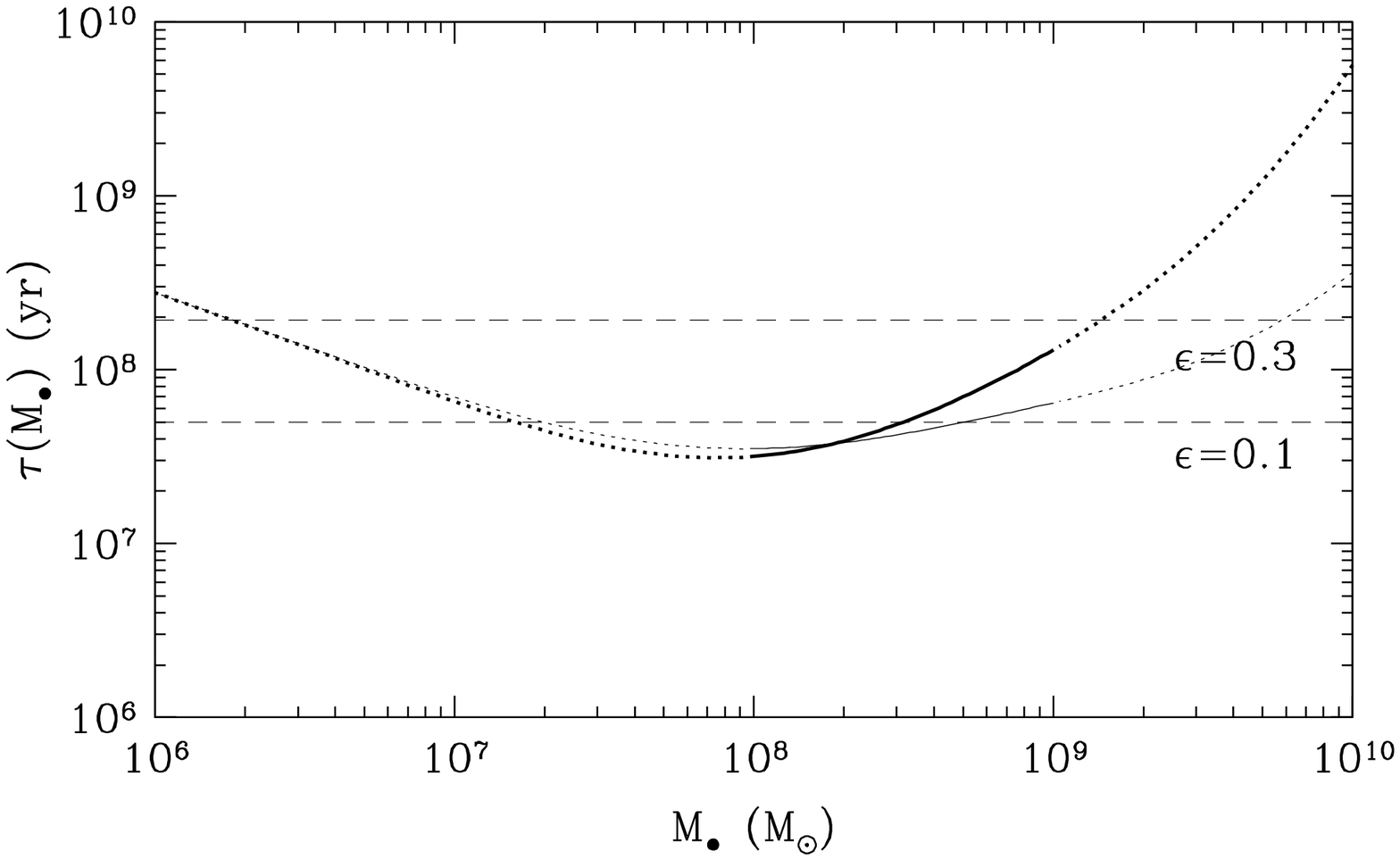}
\caption{The mean lifetime of QSOs with BH mass $\ga M\bh$
as defined by equation (\ref{eq:tau}).
The thick/thin and solid/dotted lines have similar meanings to those
in Figure~\ref{fig:eff}.
The horizontal dashed lines represent the Salpeter time (the time for a BH
radiating at the Eddington luminosity to e-fold in mass)
$\sim 4.5\times 10^7\epsilon/[0.1(1-\epsilon)]\yr$.
See \S~\ref{subsubsec:lifetime} for discussion.
}
\label{fig:lifetime}
\end{center}
\end{figure}

\noindent
If the duty cycle of QSOs with BH mass $M\bh$ is $\delta(M\bh,t)$ at time $t$,
we may define the mean lifetime of QSOs as follows:
\be
\tau(>M\bh)\equiv\frac
{\int_{M\bh}^\infty \d M\bh' \int_0^{t_0} \delta(M\bh',t)n_{M\bh}(M\bh',t)\d t}
{\int_{M\bh}^\infty n_{M\bh}(M\bh',t_0)\d M\bh'}.
\ee
With this definition, we have
\begin{eqnarray}
\tau(>M\bh)=
\frac{\int_{L(M\bh)}^\infty \d L\int_0^{t_0}\Psi(L,t)\d t}{\int_{M\bh}^\infty n_{M\bh}(M\bh',t_0)\d M\bh'}
\simeq
\frac{\int_{L(M\bh)}^\infty \d L\int_0^{t_0}\Psi(L,t)\d t}{\int_{M\bh}^\infty n_{M\bh}\early(M\bh',t_0)\d M\bh'};
\label{eq:tau}
\end{eqnarray}
the approximate equality holds for $M\bh\ga 10^8\msun$, where early-type
galaxies dominate the number density.  The lifetime $\tau(>M\bh)$ in equation
(\ref{eq:tau}) is independent of any assumed value of the efficiency
$\epsilon$.  By inserting the QSO luminosity function and the BH mass function
in local early-type galaxies in equation (\ref{eq:tau}), and assuming that
QSOs have the Eddington luminosity, we obtain the lifetime $\tau(>M\bh)$ as a
function of $M\bh$ in Figure~\ref{fig:lifetime}.  As seen from
Figure~\ref{fig:lifetime}, the mean lifetime obtained by using the
mass--dispersion relation in \citet{Tremaine02} (thick solid line) is
$\tau(>M\bh)\sim$ (3--13)$\times 10^7\yr$, i.e., the average duty cycle of
QSOs is about $\tau/t\Hubble\simeq$ 3--13$\times 10^{-3}$, where $t\Hubble$ is
the Hubble time $\sim 10^{10}\yr$.  This range is comparable to the Salpeter
time $\sim 4.5\times 10^7\epsilon/[0.1(1-\epsilon)]\yr$, the time for a BH
accreting with the Eddington luminosity to e-fold in mass (for a BH with spin
in an equilibrium state, the time for a BH radiating at the Eddington
luminosity to e-fold in entropy is half of the Salpeter time).  Hence,
accretion during bright QSO phases can significantly increase the mass of BHs,
which is once again consistent with the argument that growth of high-mass BHs
comes mainly from accretion during optically bright QSO phases.  The lifetimes
of QSOs derived here are also consistent with some theoretical models for the
luminosity function of QSOs (e.g. \citealt{HNR98,HL98}) and with the spatial
clustering of QSOs (e.g. \citealt{HH01,MW01}). 

%%%%%%%%%%%%%%%%%%%%%%%%%%%%%%%%%%%%%%%%%%%%%%%%%%%%%%%%%%%%%%%%%%%%%%%%%%%
\section{Conclusions} \label{sec:discon}

\noindent
In this paper, we have studied the observational constraints on the growth
of massive BHs in galactic nuclei.
We use the velocity dispersions of early-type galaxies obtained by the SDSS
and the relation between BH mass and velocity dispersion to
estimate the local BH mass density to be
$\rho\bh(z=0)\simeq (2.5\pm 0.4)\times 10^5 h_{0.65}^2\msun\Mpc^{-3}$
(eq.~\ref{eq:rhosigma}).
We also use the QSO luminosity function from the 2dF QSO Redshift Survey
to estimate the BH mass density accreted during optically bright QSO phases
to be
$\rho\bhaccr\QSO(z=0)=2.1\times 10^5 (C_B/11.8)[0.1(1-\epsilon)/\epsilon]\msun\Mpc^{-3}$ (eq.~\ref{eq:rhoacc0}).
These two results are consistent 
if the QSO mass-to-energy conversion efficiency $\epsilon$ is $\simeq0.1$.

By studying a continuity equation for the BH distribution and
including the effect of BH mergers in two extreme cases
(the classical case and the adiabatic case),
we predict relations between the local BH mass function and the QSO
luminosity function.
In the classical case, the predicted relation is not consistent with the
observations at high BH masses ($M\bh\ga10^8\msun$)
unless luminous QSOs ($L\bol\ga 10^{46}{~\rm erg~s^{-1}}$) have an
efficiency higher than 0.1 (e.g. $\epsilon\sim$0.2,
which is possible for thin-disk accretion onto a Kerr BH; see
Elvis, Risaliti \& Zamorani 2002 for a similar conclusion). Possible ways to
evade this conclusion include super-Eddington luminosities \citep{B01,B02} or
ejection of BHs from galactic centers, but if these are not important then 
we will come to the following conclusions.
Accretion other than what is traced by optically bright QSOs is not important
for the growth of high-mass ($\ga 10^8\msun$) BHs
(e.g. accretion by obscured QSOs, accretion
via advection dominated accretion flow, or accretion of non-baryonic
dark matter).
If the growth of low-mass BHs also occurs mainly during optically bright
QSO phases, less luminous QSOs ($L\bol\la 10^{46}{~\rm erg~s^{-1}}$)
must accrete with a low efficiency $\la 0.1$;
alternatively, other mechanisms such as obscured accretion may contribute
to the mass density of low-mass BHs, and their efficiency could be similar to
high-mass BHs, $\epsilon\sim$0.2.
The comparison of the observations with the predicted relation in the
adiabatic case also suggests that BHs in most luminous
QSOs should be rapidly rotating Kerr BHs with an efficiency $\ga 0.1$,
or there should exist more than one BH in a QSO.
We may expect that these constraints on BH demography are useful for
estimating the gravitational wave signal from BH mergers that may be detected
by {\it Laser Interferometer Space Antenna} ({\it LISA}).

We estimate the mean lifetime of luminous QSOs with BH mass in the range
$10^8\msun\la M\bh\la 10^9\msun$, which is (3--13)$\times 10^7\yr$
and comparable to the Salpeter time if $\epsilon\sim$0.1--0.3,
which is once again consistent with the argument that the growth of
high-mass BHs ($M\bh\ga 10^8\msun$) comes mainly from accretion
during optically bright QSO phases.

We are grateful to Amy Barger, Mitchell Begelman, Xiaohui Fan, Jeremy Goodman,
Martin Haehnelt, Zolt\'an Haiman, Youjun Lu, David Schlegel, David Spergel and
Michael Strauss for helpful discussions.  This research was supported in part
by NSF grant AST-9900316 and by NASA grant HST-GO-09107.09-A.

\end{document}